\documentclass [prb,preprint,superscriptaddress,floatfix] {revtex4-1}
\usepackage{amsmath}
\usepackage{graphicx}
\usepackage{lmodern}
\usepackage{amsmath}
\usepackage{hyperref}
\usepackage{color}
\usepackage{amssymb}
\usepackage{bm}
\usepackage{ulem}

\newcommand{\beq} {\begin{equation}}
\newcommand{\eeq} {\end{equation}}
\newcommand{\bea} {\begin{eqnarray}}
\newcommand{\eea} {\end{eqnarray}}
\newcommand{\be} {\begin{equation}}
\newcommand{\ee} {\end{equation}}

\DeclareMathOperator{\sgn}{sgn}

\begin{document}
\title {Eliashberg theory of phonon-mediated superconductivity -- when it is valid and how it  breaks down}
\author{Andrey V. Chubukov}
\affiliation{School of Physics and Astronomy and William I. Fine Theoretical Physics Institute,
University of Minnesota, Minneapolis, MN 55455, USA}
\author{Artem Abanov}
\affiliation{Department of Physics, Texas A\&M University, College Station,  USA}
\author{Ilya Esterlis}
\affiliation{Department of Physics, Harvard University}
\author{Steven A. Kivelson}
\affiliation{Department of Physics, Stanford University}

\date{\today}
\begin{abstract}
 We analyze the validity of Eliashberg theory of phonon-mediated superconductivity in 2D systems in light of recent extensive Monte-Carlo studies of the Holstein model.  Conventional wisdom says that Eliashberg theory is applicable as long as vertex corrections remain small. For small ratio of the phonon energy $\Omega_0$ and the Fermi energy $E_F$, this condition is supposed to hold even when the dimensionless electron-phonon coupling $\lambda$ is larger than one, i.e., in the strong coupling  regime. A comparison between various quantities computed in the Migdal approximation and those computed by Quantum Monte Carlo prove that this belief is wrong, and we identify analytically some of the ways in which this breakdown occurs for various ``normal state'' properties at $\lambda = \lambda_{cr}$, where $\lambda_{cr} = O(1)$. The breakdown
 occurs at temperatures  high enough that neither superconducting nor charge-density wave correlations extend over any significant range of distances, so it cannot be associated with the onset of an instability toward any of the relevant ordered ground-states - rather it is associated with the local physics of classical bipolaron formation.   Still, we show that certain properties, including the superconducting $T_c$ and the superconducting gap structure below $T_c$, can be accurately inferred from the strong-coupling limit of Eliashberg theory at $\lambda \leq \lambda_{cr}$.
\end{abstract}
\maketitle

\section{ Preface}

It is our great pleasure to present this mini-review for the special issue of Annals of Physics devoted to 90th birthday of Gerasim Matveevich Eliashberg.
His works on phonon-mediated suprerconductivity gave the community a much needed tool to compute $T_c$ and analyze the
properties of superconductors below $T_c$. The Eliashberg theory of superconductivity has been applied with
great success to both conventional and unconventional superconductors, and up to now remains the most reliable tool for analytical studies of superconductivity in correlated electron materials and its interplay with other effects,
including non-Fermi liquid physics.  His works form the base for our study. We send Gerasim Matveevich our very best wishes on his anniversary.

\section{Introduction}

 Electron-phonon interactions determine
many of the electronic properties of quantum materials, which include
 electrical transport
properties of most metals at all but the lowest temperatures, and the instabilities towards  superconducting
(SC) and/or charge-density-wave (CDW) states.  Phonon-mediated attractive interactions between fermions is
 the pairing glue in
the BCS theory of superconductivity.  BCS theory, however, is
 valid only at weak coupling, when the dimensionless fermion-boson coupling $\lambda$ is small. It includes only a subset of processes which give rise to logarithmically singular renormalizations of the pairing vertex at low frequencies, and approximates the full dynamical phonon-mediated interaction by a finite attraction  up to a  certain  energy cutoff, above which the interaction is set to zero.  As a result,  the pairing instability temperature $T_c$ and the gap function $\Delta (T)$ below $T_c$  depend on the cutoff; only their ratio $2\Delta (0)/T_c = 3.53$ is cutoff independent.

The Eliashberg theory (ETh) of phonon-mediated superconductivity, developed  a few years after BCS, keeps the full frequency dependence of the phonon-mediated interaction. Because the
 phonon propagator  decays at high  frequencies,  the pairing problem is ultra-violet convergent and does not need a cutoff.   Eliashberg,~\cite{Eliashberg}  and Migdal~\cite{Migdal} before him,
   argued that when the phonon  frequency $\Omega_0$ (Debye frequency  for an acoustic phonon)  is much smaller than the Fermi energy $E_F$ (i.e., the sound velocity $v_s \sim
   \Omega_0 a$ is much smaller than the Fermi velocity $v_F \sim E_F/k_F \sim E_F a$, where $a$ is the lattice constant) the corrections to the two side vertices in the pairing interaction can be neglected,  and the pairing can be analyzed by summing
   the ladder series
    in the particle-particle channel. The physical argument  underlying this observation is that in the processes leading to vertex corrections, fermions vibrate at frequencies near a bosonic mass shell,
     which  are
     thus not close to  their own  mass shell.

Due to the same smallness of
$\Omega_0/E_F$
 one can also i) neglect the  Landau damping of the phonons due to a decay into particle-hole pairs, ii) linearize the fermionic dispersion
      near $k=k_F$, and iii)
      factorize the momentum integration in each cross-section in the ladder series
       by keeping the dependence on the momentum component perpendicular to the Fermi surface only in the propagators of fast electrons and
       restricting the bosonic momenta to
       those that connect two points on the Fermi surface.
       This last consideration is relevant to cases in which the phonon propagator depends on momentum, e.g., for an acoustic phonon.

      Within these approximations, one can obtain
       a closed form integral equation relating
       the  frequency dependent dynamical gap function $\Delta (\omega, T)$
       to a convolution of $\Delta (\omega', T)/|\omega'|$ and the
          imaginary part of the effective phonon-mediated interaction, $V^{''} (\omega-\omega')$,  averaged over the Fermi surface.
     The solution of this equation for infinitesimally small
           gap function
           yields $T_c$, and the solution  for $T < T_c$ yields a finite $\Delta (\omega, T)$, which determines, e.g., the tunneling  density of states.
            (We will henceforth  incorporate the
            angle-dependent
             fermionic density of states
              at the Fermi level, $N_F$, into the definition of $V(\Omega)$, which makes it dimensionless.)
                              The
                dimensionless
          $V^{''}(\Omega)$  is commonly
          represented as $\alpha^2 F(\Omega)$, where $\alpha$  is the effective electron-phonon coupling (with units of energy) and $F(\Omega)$ is the imaginary part of a phonon propagator.  The ETh allows one to express measurable quantities in terms of $\alpha^2 F(\Omega)$, and also allows one to solve the inverse problem and extract $F(\Omega)$ from the tunneling data. An excellent agreement between the functional form of  $F(\Omega)$, extracted by Bill McMillan and
     John Rowell\cite{McMillanRowell} from the tunneling spectra in lead, and the imaginary part of the phonon propagator, inferred from inelastic neutron scattering data,
      is
      widely considered
      to be the most convincing
      single piece of evidence that  the pairing glue in a conventional superconductor is indeed
      phonon exchange.

 The frequency integral of  $\alpha^2 F (\Omega)$ determines the dimensionless coupling parameter in ETh
   \beq
   \lambda = \frac{2}{\pi} \int_{0}^\infty dx \frac{\alpha^2 F(x)}{x} = V(0)
     \ .
   \label{n_1}
\eeq
 For a single Einstein phonon with frequency $\Omega_0$,
  $V^{''}(\Omega) = \alpha^2 F(\Omega) = (\pi/2\Omega_0) \delta (\Omega - \Omega_0)$, and
  $\lambda = \alpha^2/\Omega^2_0$.

  At weak coupling, $\lambda \ll 1$, ETh reduces to
  BCS theory at frequencies $\omega, \Omega \ll \Omega_0$,
   but also allows one to accurately analyze the behavior of the system at bosonic and fermionic frequencies of order $\Omega_0$, and  to obtain $T_c$ and $\Delta (\omega, T)$ for a given $\alpha^2 F(\Omega)$. It
   has been argued, however~\cite{Carbotte_90,Marsiglio_91,Karakozov_91,combescot,Wang2016}
    that ETh remains valid
    even when $\lambda$ becomes larger than
   1, i.e., at strong coupling.  The argument, due to Migdal~\cite{Migdal} and Eliashberg~\cite{Eliashberg}
    is that the small parameter, which allows one to neglect vertex corrections, is of order
   $\lambda v_s/v_F \sim \lambda \Omega_0/E_F$. For $\Omega_0 \ll E_F$, this  parameter remains small even when $\lambda >1$, up to $\lambda \sim E_F/\Omega_0$.

    At strong coupling, ETh has to
    take into account the fermionic self-energy $\Sigma (k, \omega)$ as the strength of the self-energy corrections
    to the electron propagator are controlled by $\lambda$.
      For the calculations of $\Sigma (k, \omega)$,
      the same line of reasoning suggests that vertex corrections again can be neglected, and
     the momentum integration can be factorized.  As
      a consequence, the self-energy depends on frequency
     more strongly  than on momentum and can be approximated by $\Sigma (\omega)$.  The equations for $\Delta (\omega)$ and $\Sigma (\omega)$ form a coupled set:
     $T_c$ and the form of the gap function below $T_c$ are affected by the self-energy, and the self-energy in turn gets modified below $T_c$.

    The strong coupling limit of ETh attracted
  considerable  attention in the past because in this limit the solution of the Eliashberg equations yields~\cite{ad,Carbotte_90}
    $T_c =0.1827 \Omega_0 \sqrt{\lambda} =0.1827 \alpha$, which is much larger than $\Omega_0$, and
    also
    because the forms of $\Delta (\omega)$ and of the tunneling density of states are highly non-trivial~\cite{combescot} (see Sec. \ref{ssec:strong}).
    That the onset temperature of the pairing is parametrically larger than $\Omega_0$ is puzzling at
     first glance  because
     the
     phonon-mediated
      interaction $V(\Omega) \propto 1/(\Omega^2 - \Omega^2_0)$ is attractive up to $\Omega_0$ and repulsive at higher frequencies, and at $\Omega_0 \to 0$ the region of attraction shrinks.
  It was argued~\cite{combescot}
    that although $T_c$ remains finite, the pairing problem at strong coupling is very different from BCS and
     can be effectively described as self-trapping, i.e., a process in which if a system develops a pairing gap,  the pairing potential gets modified in such a way that it favors a larger gap.
   The authors of another paper in this volume~\cite{Chubukov_2019_this_volume} argued that in this situation
     ETh corresponds to a shallow minimum of the Free energy, i.e., fluctuations  beyond ETh are strong, despite that Eliashberg $T_c$ is much smaller than $E_F$.  It remains to be seen how much these fluctuations reduce $T_c$ down from its mean-field value.

The relation between superconducting $T_c$ and  the energy of a soft boson  has been extensively discussed for pairing near a quantum critical point (QCP) in a metal.  There, the pairing is mediated by
 a soft collective boson, which represents the fluctuations of
 a
 spin or charge  order parameter that condenses at the QCP.  A finite $T_c$ at a critical point
 suggests the existence of a dome of superconductivity above a QCP,
 similar to what
  has been observed in several classes of materials.  However, for electron-phonon superconductors, there is no experimental evidence so far that $T_c$
  ever exceeds
  (or even comes close to equalling)
   $\Omega_0$. Furthermore, recent extensive Determinant Quantum Monte Carlo (DQMC) calculations for the Holstein model,
   \cite{QMC,DQMC} which is the paradigmatic model for phonon-mediated superconductivity,  have found that $T_c$ is at most $0.1\ \Omega_0$ even for the case when $\Omega_0$ is much smaller than $E_F$ and vertex corrections, which could potentially
   lead to a breakdown of the ETh
   should be small.

   In this communication we discuss
  the origin of the apparent discrepancy between DQMC data and the strong coupling limit of the ETh.
  At the most basic level, the ``bare" ETh
  breaks down at $\lambda = O(1)$  due to the renormalization
   of the static phonon propagator by the fermionic  polarization bubble.  The strength of this renormalization is determined by  $\lambda$ rather than by $\lambda \Omega_0/E_F$.
     For instance, in the rotationally-invariant case, the one-loop renormalization  does not depend on bosonic momentum and changes $\Omega_0$ into $\Omega^{eff}_0 = \Omega_0 (1 - 2 \lambda)^{1/2}$ (see Sec. \ref{sec:Rbosonic} below).  The ETh is then only valid
     at most
      up to $\lambda =1/2$ and
      the
      maximum possible $T_c$ remains a fraction of $\Omega_0$.
     Still, near $\lambda =1/2$, one can construct
     an  effective ETh with the bosonic propagator with $\Omega^{eff}_0$ instead of $\Omega_0$ and with the new coupling  $\lambda^{eff} =  \lambda/(1-2 \lambda)$.  This effective ETh is in the strong coupling limit for $\lambda \leq 1/2$, and the corresponding $T_c$ behaves as  $T_c \approx 0.1827 \Omega^{eff}_0 \sqrt{\lambda^{eff}}$.  This $T_c$ is parametrically larger than $\Omega^{eff}_0$, but is still a fraction of the bare $\Omega_0$.
     While,  as discussed in the next paragraph, various normal state properties are  not well represented even in this ``effective'' sense,  for the purposes of determining specific properties of the superconducting state  ETh near $\lambda = 1/2$ {\em may} be valid in 2D as long as
  $ \lambda^{eff}\ ( \Omega_0/E_F) \  S_{2D}$ remains small, where $S_{2D} = \pi \log{E_F/\Omega_0}$
    is a logarithmic  factor specific to 2D (see Sec. \ref{sec:Rbosonic}).
    For a lattice system, the renormalization does depend on momentum and changes $\Omega_0$ into
   $\Omega^*_0 (q)$.  For the dispersion used in the DQMC study, the renormalization of $\Omega_0$ by the fermionic polarization bubble is strongest at $q_0=(\pi,\pi)$.
     In this case, $\Omega^{eff} _0 (q)$ has a minimum at $q=q_0$.
     For $\lambda > \lambda_{cr}$, the system develops $(\pi,\pi)$ CDW order at low $T$.
       (The $T=0$ transition to the CDW state appears to be first order, so while the softening of $\Omega^{eff}_0(\pi,\pi)$ is substantial, it is never seen to go strictly to zero.)
For $\lambda \lesssim \lambda_{cr}$ one can
 construct an effective model near $\lambda = \lambda_{cr}$ with a $q-$dependent bosonic propagator, and study it within the ETh.
       The corresponding $T_c$
       exceeds $\Omega^{eff}_0 (q_0)$, but
      remains small compared to both the bare $\Omega_0$ and to $\Omega^{eff}_0 (q)$, averaged over the Fermi surface.

We also analyze the phase diagram at $T> T_c$. DQMC results show\cite{QMC,DQMC} that there exists a crossover line $\lambda= \lambda_{cr} (T)$, which
  separates a  Fermi liquid  at smaller $\lambda$
 from
 a classical bipolaron lattice gas at larger $\lambda$.
 This is an additional way in which corrections to Migdal theory (vertex corrections) alter the physics at large $\lambda$, although in a way that has relatively less impact on the superconducting state itself.
  We speculate that this crossover may be associated with singular thermal contribution to the self-energy $\Sigma (\omega) =  i T \lambda^{eff}\mbox{sign}[ \omega]$. This thermal self-energy  acts as a non-magnetic impurity and cancels out in the gap equation, but does give rise to
  precursors of a bipolaron gas, much like thermal spin fluctuations give rise to thermal precursors to  a SDW state.

The paper is organized as follows. In Sec. \ref{sec:eli} we  briefly summarize the original ETh of  electron-phonon superconductivity, introduce the effective coupling $\alpha$,  and discuss
the
 weak and strong coupling regimes. In Sec. \ref{sec:beyond} we analyze the validity of ETh in 2D. We obtain
an explicit expression for the vertex correction and show that in 2D there is an additional logarithm, not present in 3D. We then discuss the corrections to
the bosonic propagator. In Sec. \ref{sec:effective} we discuss
the effective ETh with renormalized  $\Omega^{eff}_0$ and $\lambda^{eff}$ for both rotationally invariant and lattice systems and analyze the crossover induced by thermal corrections to the fermionic self-energy.
    In Sec. \ref{sec:QMC} we introduce the Holstein model and discuss results of the
 DQMC  analysis \cite{QMC,DQMC}. We summarize our results in Sec. \ref{sec:conclusions} and discuss our findings  in a broader context in Sec. \ref{sec:conclusions_1}

\section{
Eliashberg theory of
phonon-mediated superconductivity}
\label{sec:eli}

We begin with a brief review of the canonical ETh of
 phonon-mediated superconductivity.
As our purpose is to discuss the limits of validity of ETh we avoid unnecessary complications and consider a simple model of
 fermions with parabolic dispersion coupled  to an Einstein phonon.
We consider only electron-phonon interactions,
{\it i.e.} we
 neglect direct Coulomb repulsion between the fermions.
 The analysis of the interplay between Coulomb repulsion and electron-phonon interaction is rather involved and requires
 separate considerations. (See the article by Ruhman et al in this volume.)

Exchange of an Einstein phonon gives rise to an effective 4-fermion interaction
\beq
V(\Omega) = \frac{\alpha^2}{\Omega^2_0 - (\Omega + i \delta)^2} \ .
\label{aa_1}
\eeq
Here, we incorporate the fermionic density of states $N_F$ into the definition of $\alpha$, so that
 $\alpha$ has
the dimensions of energy and $V(\Omega)$ is dimensionless.
   The effective  interaction $V(\Omega)$ causes renormalizations in both
   the particle-hole and particle-particle channels.
   In the particle-hole channel, $V(\Omega)$  gives rise to
   a dynamical fermionic self-energy,
   that makes
   the fermions less coherent. In the particle-particle channel,
   the same $V(\Omega)$
   gives rise to pairing below a certain $T$.
  The two effects are treated on equal footings in the ETh, i.e., the tendency to pairing is affected by
  the fermionic self-energy, while the fermionic self-energy changes when the system
    becomes a superconductor.

   As we said in the Introduction,   ETh neglects corrections to
   the fermion-boson vertex from processes involving particle-hole bubbles.
   Consequently, the fermionic self-energy is computed self-consistently within one-loop approximation, but with
   the full normal and anomalous Green's functions, and the pairing vertex is computed within the ladder approximation, again with the full Green's functions.
     In particular, ETh neglects Kohn-Luttinger corrections to the pairing vertex.  The ETh also  assumes that pairing involves fermions with energies much smaller than $E_F$ and thus uses
    the fermionic dispersion linearized near the Fermi surface.
    \footnote{In this last respect ETh is qualitatively different from dynamical mean-field theory, for which
    the finite bandwidth of the fermions is a necessary ingredient.}

   Within these approximations one can obtain a closed set of coupled integral equations for two frequency dependent functions  -- the fermionic self-energy $\Sigma (\omega)$ and the pairing vertex $\Phi (\omega)$. The pairing vertex is generally a function of both bosonic and fermionic frequencies, $\Phi (\omega, \Omega - \omega)$. In ETh it is taken at the bosonic $\Omega =0$
    and is a function of a running fermionic frequency $\omega$.  Below we will use ${\tilde \Sigma} (\omega) = \omega + \Sigma (\omega)$.

   Eliashberg equations are most commonly analyzed on the  Matsubara axis, where $\omega_m$ form a discrete set $\omega_m  = \pi T (2m +1)$.  Here, the two equations are
     \bea
   &&\Phi(\omega_m) = \alpha^2 \pi T \sum_{\omega_{m'}} \frac{\Phi (\omega_{m'})}{\sqrt{{\tilde \Sigma}^2 (\omega_{m'}) + \Phi^2 (\omega_{m'})}} \frac{1}{(\omega_m - \omega_{m'})^2 + \Omega^2_0} \nonumber \\
     && {\tilde \Sigma} (\omega_m) = \omega_m + \alpha^2  \pi T \sum_{\omega_{m'}} \frac{{\tilde \Sigma} (\omega_{m'})}{\sqrt{{\tilde \Sigma}^2 (\omega_{m'}) + \Phi^2 (\omega_{m'})}} \frac{1}{(\omega_m - \omega_{m'})^2 + \Omega^2_0} \ .
   \label{2}
   \eea
  Even for $T<T_c$, a reference  ``normal state'' solution to these equation can be obtained by setting $\Phi=0$. In
   such normal state
    at $T=0$,
    \beq
    \Sigma (\omega_m) =  \lambda  \Omega_0 \arctan{\frac{\omega_m}{\Omega_0}}
    \label{nn_1}
    \eeq
 with $\lambda = V(0) =\alpha^2/\Omega^2_0$.
  At a finite $T$,
     \beq
    \Sigma (\omega_m) =  \lambda  \pi T \left(1 + 2 \left(\frac{\Omega_0}{2\pi T}\right)^2 \sum_{1}^m
     \frac{1}{n^2 + \left(\frac{\Omega_0}{2\pi T}\right)^2}\right)
    \label{nn_2}
    \eeq
     for $m >0$, and $\Sigma (\omega_{-(m+1)}) = - \Sigma (\omega_m)$. At the first two Matsubara frequencies, $m =0$ and $m =-1$ ($\omega_m = \pm \pi T$) the second term in the r.h.s. of
     (\ref{nn_2}) vanishes,
     such that
     \beq
     \Sigma (\pm \pi T) = \pm \pi T \lambda.
     \label{nn_3}
    \eeq
\begin{figure}
\includegraphics[width=.8\columnwidth]{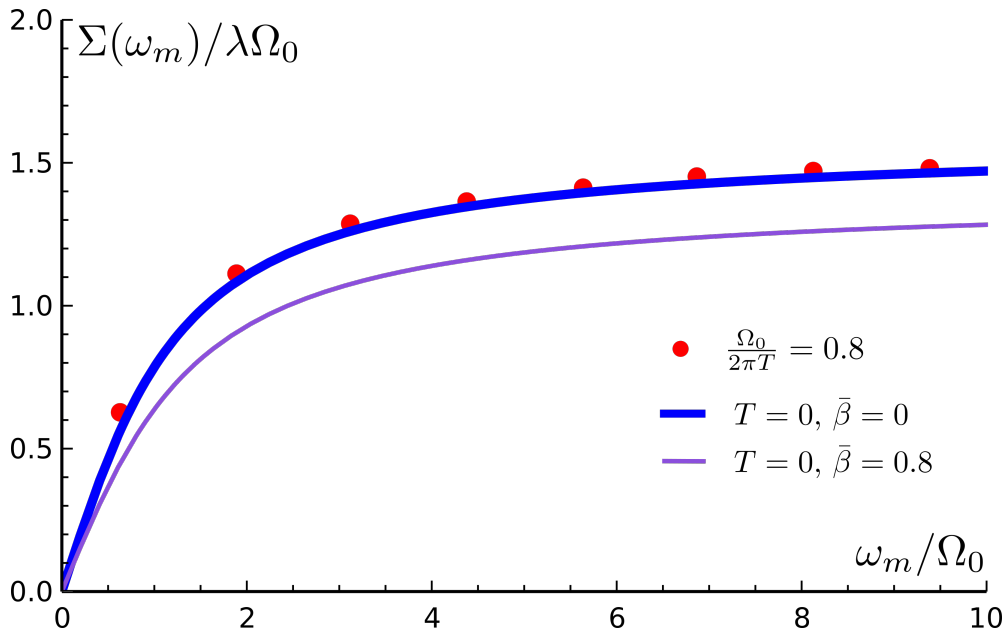}
\caption{Fermionic self-energy $\Sigma (\omega_m)$ in the normal state. The self-energy is linear in frequency at
 small $\omega_m$ and saturates at a finite value at large $\omega_m$. The canonical ETh is the case ${\bar \beta} =0$.  The self-energy in the ETh is essentially independent on $T$, as evidenced by near-equivalence of the results at $T=0$ and at $T = \Omega_0/(1.6 \pi)$.
  The curve for a non-zero  ${\bar \beta}$ is for the case when the  bosonic propagator has momentum dependence, induced by dressing the propagator by fermionic particle-hole bubbles (see Eq. \ref{nn_3_1}).
We used  $\lambda =0.3$, $\Omega_0/E_F = 0.08$.}
 \label{fig:self_energy}
\end{figure}
 We plot the self-energy in the normal state at $T=0$ and at finite $T$ in Fig. \ref{fig:self_energy}.

 The Eliashberg equations can be conveniently re-arranged by introducing the gap function $\Delta (\omega_m)$ and the function $Z(\omega_m)$  via
    \beq
   \Delta (\omega_m) = \Phi (\omega_m) \frac{\omega_m}{{\tilde \Sigma} (\omega_m)}
   \label{3}
   \eeq
    and
    \beq
    Z(\omega_{m}) = \frac{{\tilde \Sigma} (\omega_m)}{\omega_m}
    \label{4}
    \eeq
 At vanishing $T$ and in the limit $\omega_m \to 0$,  the function $Z (0)$ -- the ``Eliashberg $Z$-factor,'' is the inverse of the quasiparticle residue $Z_{QP} = 1/Z(0)$.

   In terms of these new functions $Z(\omega_{m} )$ and $\Delta (\omega_{m})$, the Eliashberg equations become
    \bea
   &&\Delta (\omega_m) = \alpha^2  \pi T \sum_{\omega_{m'}}\frac{1}{\sqrt{\omega^2_{m'} +
   \Delta^2 (\omega_{m'})}}    \left(\Delta (\omega_{m'}) - \Delta (\omega_m) \frac{\omega_{m'}}{\omega_m} \right) \frac{1}{(\omega_m - \omega_{m'})^2 + \Omega^2_0} \label{5}\\
   \label{gapequation}
   && Z(\omega_m) = 1 + \frac{\alpha^2}{\omega_m}  \pi T \sum_{\omega_{m'}} \frac{\omega_{m'}}{\sqrt{
   \omega^2_{m'} + \Delta^2 (\omega_{m'})}}
    \frac{1}{(\omega_m - \omega_{m'})^2 + \Omega^2_0} \ .
   \label{6}
   \eea
  The advantage of presenting the equations in this form is that
  Eq.  \eqref{5} for $\Delta (\omega_m)$ does not depend on $Z(\omega_m)$ and
  Eq. \eqref{6} for $Z(\omega_m)$ depends only on $\Delta (\omega_{m'})$.
 Hence
  one first solves for  $\Delta (\omega_m)$ and then uses it to obtain $Z(\omega_m)$.
 The
lack of any explicit $Z$ dependence of the gap equation, Eq. \ref{gapequation},
 reflects the fact that the objects that undergo pairing are quasiparticles, whose distribution function does not
 depend on the residue $Z_{QP}$.

In the normal state at $T=0$, the self-energy is linear in $\omega_m$ at small frequencies, $\Sigma (\omega_m) = \lambda \omega_m$. In this limit, $Z (\omega_m) = 1 + \lambda$ coincides with the inverse residue of the fermionic propagator $G_k (\omega_m) = Z^{-1}/(i \omega_m - v^*_F (k-k_F))$, where $v^*_F = v_F/Z$.

  Within ETh, one can also compute the Free energy in the superconducting and the normal state, $F_{sc}$ and $F_n$, and the mean-field condensation energy $\delta F = F_{sc} - F_n$.
 The condensation energy $\delta F$ depends only on $\Delta (\omega_m)$ (Refs.~\cite{lw,Bardeen,Eliashberg}):
 \bea
 \delta F  &=& - 2\pi T N_F \sum_m  |\omega_m| \left(\frac{1}{\sqrt{1 + D^2_m}}-1\right) \nonumber \\
    && -  \pi^2 T^2 \alpha^2  \sum_{m, m'} \frac{\sgn \omega_m \sgn \omega_{m'}}{|\omega_m-\omega_{m'}|^2 + \Omega^2_0} \frac{1 + D_m D_{m'} - \sqrt{1 + D^2_m} \sqrt{1 + D^2_{m'}}}
    {\sqrt{1 + D^2_m } \sqrt{1 + D^2_{m'}}}
\label{7}
 \eea
where $D_m = \Delta (\omega_m)/\omega_m$.
  The gap equation (\ref{5}) is obtained from $\partial \delta F/\partial \Delta_m =0$.
At  $T=0$, $\delta F$ is the condensation energy of an Eliashberg superconductor.

We note that $F_{sc}$ and  $F_n$ are
 not the full Luttinger-Ward Free energies as the Eliashberg equations are obtained by minimizing Luttinger-Ward functional
 with respect to variations of $\Sigma$ and $\Phi$ (or $\Delta$ and $Z$).  Accordingly, the Eliashberg Free energies are computed right at the minimum, without fluctuation corrections and in these respect are mean-field Free energies.  The same is true for $\delta F$.

   Eqs. (\ref{2}), (\ref{5}), and (\ref{7}) can be simplified even further, by subtracting the contribution from thermal fluctuations, i.e., the term with $m'=m$ in the r.h.s. of
   the Eliashberg equations. For the equation
   for  $\Delta$ this is obvious because $\Delta (\omega_{m'}) - \Delta (\omega_m) \omega_{m'}/\omega_m$ in the numerator vanishes at $m =m'$. The same is true for Eq. (\ref{7}).   In Eqs. (\ref{2}),  one can   pull out the term with $m'=m$ from the r.h.s, move it to the l.h.s, and introduce new variables
 $\Phi^* (\omega_m)$ and ${\tilde \Sigma}^* (\omega_m)$ via
   \bea
   \Phi^* (\omega_m) &=& \Phi (\omega_m)
   \left(1- Q (\omega_m)\right), \nonumber \\
   {\tilde \Sigma}^* (\omega_m) &=& {\tilde \Sigma} (\omega_m) \left(1- Q (\omega_m)\right)
   \label{8}
   \eea
    where
    \beq
     Q (\omega_m) =   \frac{\pi T \lambda}{\sqrt{{\tilde \Sigma}^2 (\omega_{m}) +\Phi^2 (\omega_{m})}}
   \label{9}
   \eeq
   and
   \beq
   \lambda = \frac{\alpha^2}{\Omega^2_0}
   \label{10}
   \eeq
    The ratio $\Phi (\omega_m)/ {\tilde \Sigma} (\omega_m) = \Phi^* (\omega_m)/ {\tilde \Sigma}^* (\omega_m)$,
      hence the equations for $\Phi^* (\omega_m)$ and ${\tilde \Sigma}^* (\omega_m)$ are the same as for $\Phi (\omega_m)$ and ${\tilde \Sigma} (\omega_m)$, but the summation in the r.h.s. now runs over $m \neq m'$. The physical reasoning for the cancellation of the contributions from thermal phonons in Eliashberg equations
 is that thermal phonons scatter with zero frequency transfer and arbitrary momentum transfer and in this respect
  act in the same way as impurities. For $s$-wave, spin-singlet pairing, thermal phonons give equal contributions to the self-energy and the pairing vertex and mimic non-magnetic impurities.
    From this perspective, the cancellation of the thermal contribution is the manifestation of
    Anderson's theorem.
Note, however, that the thermal contribution does not cancel in $Z(\omega)$, i.e.,
  the full self-energy $\Sigma (\omega)$ does contain contributions from thermal fluctuations.

We also see from Eqs. (\ref{2}), (\ref{5}), and (\ref{8}) that the bosonic $\Omega_0$
 factors out from the summand, once we rescale the temperature $T$ to dimensionless ${\bar T} = T/\Omega_0$, and the dimensionless $\lambda$  remains the only parameter in the gap equation.  Obviously then
ETh yields an expression for the critical temperature of the  form $T_c^{(ETh)} = \Omega_0 f_F(\lambda)$. We will call this
 the ETh value of $T_c$;  it
  may better be thought of
   as the onset temperature for the pairing keeping in mind that the actual $T_c$ may be smaller because of pairing  fluctuations.

The
Eliashberg equations on the Matsubara axis can be used to obtain $T_c^{(ETh)}$ and thermodynamic properties below $T_c$, e.g., the jump of the specific heat at $T_c$.   To obtain transport properties of a superconductor one needs $\Delta (\omega)$ along the real frequency axis.  The transformation
 cannot be done by just a rotation from $i\omega_m$ by $\omega$, because
 in the complex  frequency  plane $(i \omega_m \to z)$, $ V(\omega_{m'} +iz)$
   has poles at $z = i \omega_{m'} \pm \Omega_0$.
 One needs to add additional terms to the r.h.s. of the Eliashberg  equation for the retarded $\Delta (\omega)$ to cancel these singularities and restore analyticity~\cite{combescot,Marsiglio_88,Marsiglio_91,Karakozov_91}.  Alternatively, one can use
 the  spectral representation
 to derive the Eliashberg equation for the gap function directly
 for real frequencies~\cite{Karakozov_91}. The resulting equation for $\Delta (\omega)$ has the form
\begin{equation}
D(\omega) B(\omega) = A(\omega) + C(\omega)
\label{11}
\end{equation}
where
$D(\omega) = \Delta (\omega)/\omega$ and
\begin{eqnarray}
&&A(\omega) = \frac{\alpha^2}{2} \int_0^{\infty} d \omega^\prime
\tanh{\frac{\omega^\prime}{2T}}
 \Re\left[\frac{D(\omega^\prime)}{\sqrt{1 - D^2(\omega^\prime)}} \left(\frac{1}{\Omega^2_0 -(\omega - \omega^\prime)^2} + \frac{1}{\Omega^2_0 -(\omega + \omega^\prime)^2}\right)\right] \nonumber \\
&& B(\omega) = \omega + \frac{\lambda}{2} \int_0^{\infty} d \omega^\prime
\tanh{\frac{\omega^\prime}{2T}} \Re\left[\frac{\omega^\prime}{\sqrt{1 - D^2(\omega^\prime)}}
\left(\frac{1}{\Omega^2_0 -(\omega - \omega^\prime)^2} - \frac{1}{\Omega^2_0 -(\omega + \omega^\prime)^2}\right)\right]\nonumber \\
 && C(\omega) = i \frac{\alpha^2}{2}~\int_{-\infty}^{\infty}
d\Omega \Im \frac{1}{\Omega^2_0 -(\Omega + i \delta)^2}
\left[\coth \frac{\Omega}{2T} - \tanh \frac{\Omega + \omega}{2T}\right]
 \frac{D(\omega + \Omega) -D(\omega)}{\sqrt{1 - D^2(\omega +\Omega)}} \nonumber \\
&& =  i \frac{\pi \alpha^2}{4 \Omega_0} \Bigg[ \left(\coth \frac{\Omega_0}{2T} - \tanh \frac{\omega + \Omega_0}{2T}\right)
 \frac{D(\omega - \Omega_0) -D(\omega)}{\sqrt{1 - D^2(\omega +\Omega_0)}}
\label{12} \\
 &&
+ \left(\coth \frac{\Omega_0}{2T} + \tanh \frac{\omega - \Omega_0}{2T}\right)
 \frac{D(\omega - \Omega_0) -D(\omega)}{\sqrt{1 - D^2(\omega -\Omega_0)}}\Bigg] \ . \nonumber
\end{eqnarray}
Here the integrals are the principal values.
For practical purposes, it is sometimes advantageous to use a mixed approach:
obtain the integral equation for $\Delta (\omega)$ with $\Delta (\omega_m)$ in the input term, solve for $\Delta (\omega_m)$ and find the input, and then solve for
 $\Delta (\omega)$ (Refs. \cite{combescot,Marsiglio_88,Marsiglio_91,wu_1}).

We now briefly review the solution of the Eliashberg equations.

\subsection{Weak coupling, $\lambda \ll 1$}

 At weak coupling, the solution of the Eliashberg gap equation reproduces the known results of  BCS theory:
  $T_c$ scales as $e^{-1/\lambda}$, $\Delta (\omega) \approx \Delta$
  for  $\omega < \Omega_0$,
   and $2\Delta/T_c \approx 3.53$.  The only substantial difference between the Eliashberg and BCS theories at weak coupling is that the latter requires a high-energy cutoff, which sets the pre-exponential factors for $T_c$ and $\Delta$, while in ETh the cutoff is effectively provided by the frequency dependence of $V (\Omega)$.  As
   a consequence, both $T_c$ and $\Delta$  are obtained within ETh
    with the exact prefactors,
    as has been discussed in several papers, using different computational tools~\cite{prefac1,prefac2,prefac3,prefac4,prefac5,combescot1990critical,Chubukov2016,Marsiglio18,Phan2019}.
   The result is
   \beq
   T_c =  1.13\ e^{-1/2} \Omega_0 e^{-\frac{1+\lambda }{\lambda}} = 0.252\ \Omega_0 e^{-\frac{1}{\lambda}}
   \label{14}
   \eeq
   A recipe
   for computing the weak coupling $T_c$ for an arbitrary non-critical
bosonic propagator has been given in ~\cite{combescot1990critical}.
 The gap function $\Delta (\omega)$ is a frequency independent constant, $\Delta(\omega)  =1.76\  T_c$
 for  $\omega \ll \Omega_0$, and decays as $1/\omega^2$
 for $\omega \gg \Omega_0$.

 \subsection{Strong coupling, $\lambda \gg 1$}
\label{ssec:strong}

We discuss the applicability of the strong coupling limit of ETh later in the paper. Here we just analyze Eqs. (\ref{2}) and (\ref{5}) in
 the large $\lambda$ limit which we approach by holding $\alpha$ fixed and taking $\Omega_0 \to 0$ (see Eq. \eqref{10}). Note that we define the canonical ETh as the one for which the phonon propagator is treated as given, i.e.,  does not include the  renormalization of $V(\Omega)$  by fermions.
  We will discuss this renormalization later, when we analyze the corrections to the canonical ETh.

To obtain $T_c$, we set $\Delta (\omega_m)$
 to be infinitesimally small.
 A quick look at Eq. \eqref{5} shows that the r.h.s. of the gap equation is non-singular at $\Omega_0 =0$:
 \beq
 \Delta (\omega_m) =
 \left(\frac\alpha{2\pi T_c}\right)^2 \sum_{m' \neq m }\frac{ \left[\Delta (\omega_{m'})\left({2m+1}\right) - \Delta (\omega_m) \left({2m'+1}\right)\right]}{{|2m'+1|}{(2m+1)}(m-m')^2} \ .
\label{15}
\eeq
 This equation has one dimensionless parameter  $\alpha/(2\pi T_c)$. (
Recall that
 $\alpha$ has
 the dimensions of  energy.)  Hence, if a solution exists, $T_c$ must be of order $\alpha$.  Eq.
 \ref{15} has been solved
 numerically on a large mesh of Matsubara frequencies~\cite{ad,Bergmann_1973,combescot},
 with  the result
 \beq
 T_c \approx 0.1827\ \alpha \ .
 \label{16}
 \eeq
One can analyze extensions of (\ref{15})
for the case in which instead of  $V(\Omega) = \alpha^2/\Omega^2$
we have $V(\Omega)= \alpha^\gamma/|\Omega|^\gamma$;
the resulting equations can be solved analytically in the limit of large $\gamma$, from which it follows that
 $ T_{c} = \frac\alpha{2\pi} s^{1/\gamma}$ (Ref. \cite{Wang2016}),
   where $s$ is determined from $J_{3/2+1s} (1/s)/J_{1/2+1/s} (1/s) = s-1$, and $J_a (b)$ is a Bessel function.
  The solution is $s\approx 1.1843$. Applying this  to $\gamma =2$, we obtain
  $T_c \approx 0.17 \alpha$, in good agreement with the numerical result.  We also note that $T_c$ is reasonably close
   to $\alpha/2\pi \approx 0.16 \alpha$. The same result for $T_c$ can be obtained by solving the set of equations for the pairing vertex and the self-energy.
   Note that the
    full self-energy $\Sigma (\omega_m)$ diverges at $\Omega_0 \to 0$ because of singular contributions from thermal fluctuations. However, the truncated $\Sigma^* (\omega_m)$ is free from singularities. Evaluating
    ${\tilde \Sigma}^* (\omega_m)$, substituting it into the equation for $\Phi^* (\omega_m)$, and solving the latter as an eigenvalue problem, one reproduces $T_c$ from (\ref{16}).

     Eq.  (\ref{16}) was first obtained in Ref. \cite{ad}. These authors expressed the critical temperature  as $T_c \sim \Omega_0 \sqrt{\lambda}$ to emphasize that at strong coupling, $T_c$ is parametrically larger than $\Omega_0$. Using $\lambda = \alpha^2/\Omega^2_0$, one immediately finds that
     this is equivalent to $T_c \sim \alpha$, as in (\ref{16}).

The gap function $\Delta (\omega_m)$ at $T \ll T_c$ has a universal form $\Delta (\omega_m) = \Delta (\pi T) f (\omega_m/\Delta (\pi T))$, where $\Delta (\pi T) \sim \alpha$ and $f(x \ll 1) \approx 1$
 and $f (x \gg 1) \propto 1/x^2$.
  Still, the frequency
  dependence  of $\Delta (\omega_m)$
 is stronger than in the weak coupling limit. For example,  at $T \ll T_c$, $\Delta (\pi T) \approx \Delta (0)$ is roughly $1.6$ times larger than $\Delta (\omega_m)$ at
 the frequency
 at which $\Delta (\omega_m) = \omega_m$.  The ratio of $2\Delta (0)/T_c$ is a pure number,
 as at weak coupling, but its value is close to 13, i.e., is much higher than at weak coupling. A large $2 \Delta/T_c$ ratio can be understood by again looking at the extension to  $\gamma>2$: $T_c$ saturates at $\alpha/2\pi$ at larger $\gamma$, while
 $\Delta (0)$ diverges for $\gamma =3$,  as at this $\gamma$  the singularity of the denominator in the r.h.s. of the gap equation, (\ref{5}), at $\omega_m = \omega_{m'}$  is no longer compensated by the vanishing of the numerator.
 The large value of $2\Delta/T_c$ for $\gamma =2$ (our case) reflects the fact that  for this $\gamma$ $\Delta (0)$ is already large.

Although $T_c$ is finite in the strong coupling limit of ETh and $\Delta (\omega_m)$ is a regular function  of frequency, the behavior of the gap function and the density of states
analytically continued to real frequencies is highly non-trivial~\cite{combescot,Marsiglio_88,Marsiglio_91}.
For instance, at $T=0$, the
 gap $\Delta (\omega)$  behaves as $\Delta (\omega)  \approx \omega/\sin(\phi (\omega/\Delta (0)))$, where $\phi (x)$ is a near-linear  function of the argument. At small $x$,
  $\phi (x) \approx x$ and $\Delta (\omega)  \approx \Delta (0)$, as expected, but at larger $\omega$,
   $\Delta (\omega)$ oscillates in sign and diverges at a discrete set of $\omega$ (see the left panel on Fig.\ref{fig:eli})
  [Along Matsubara axis, $\omega/\sin(\omega/\Delta (0))$ becomes $\omega_m/\sinh(\omega_m/\Delta (0))$, which is a regular function of $\omega_m$]. This behavior has been analyzed in detail in Ref. \cite{combescot}. (See also the paper by D. Hauck et al in this volume.)

\subsection{Intermediate coupling}

\begin{figure}
\includegraphics[width=.4\columnwidth]{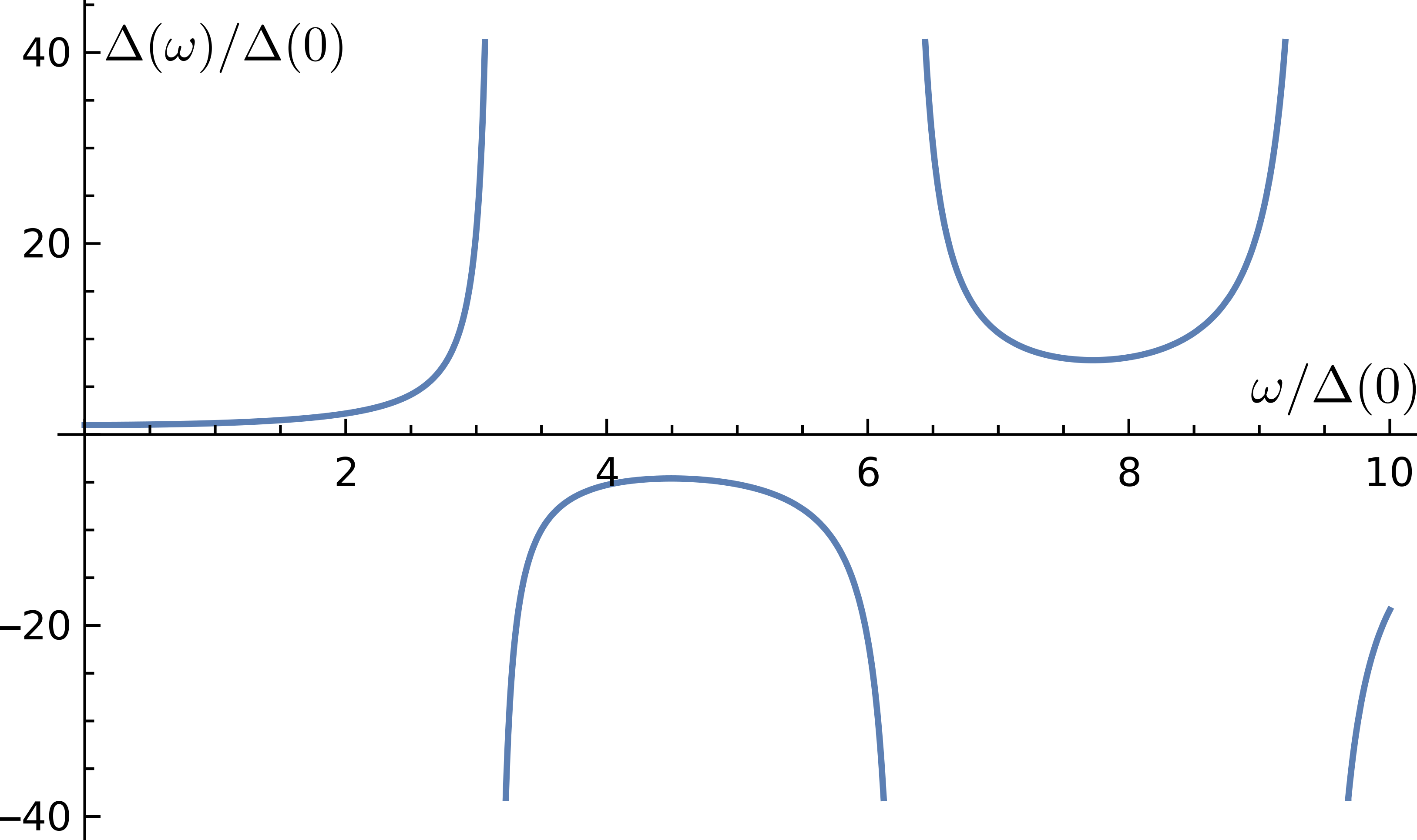}
\includegraphics[width=.4\columnwidth]{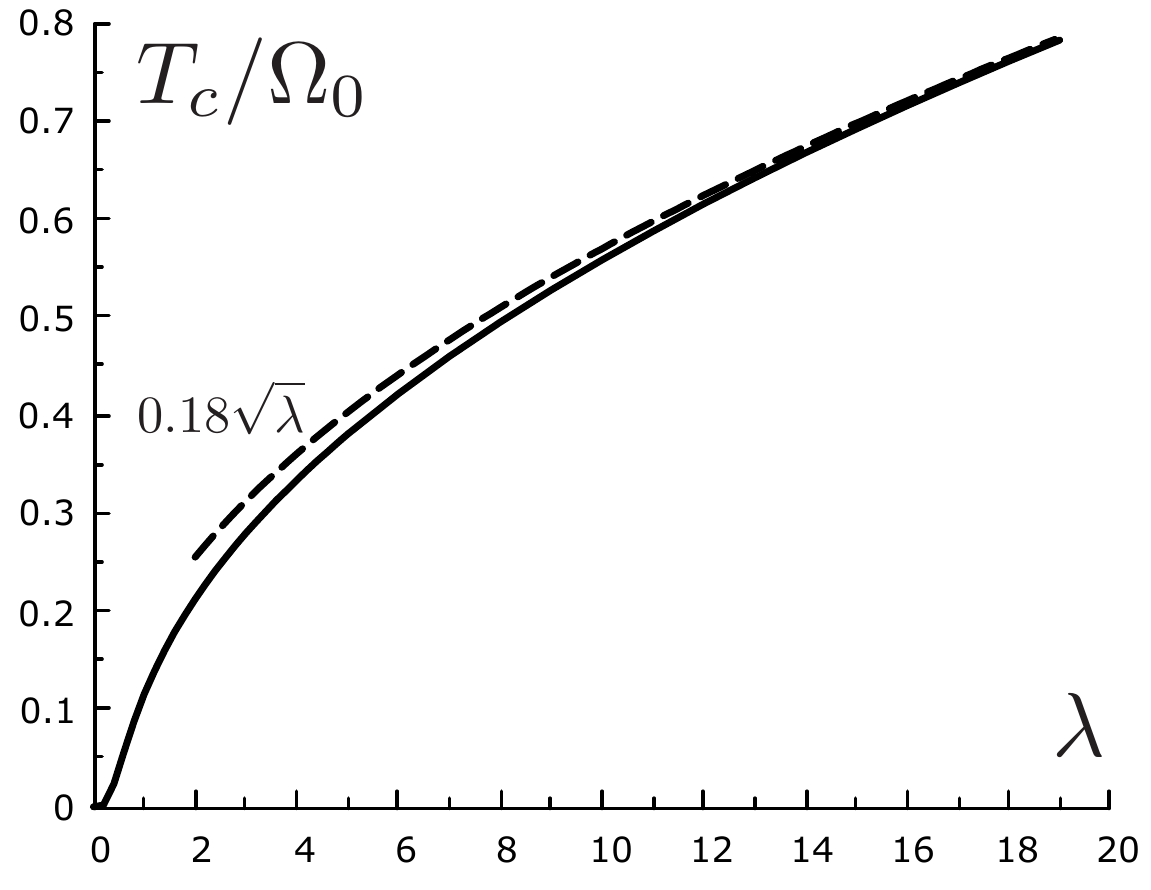}
\caption{
Left:
the gap function $\Delta (\omega )$ in real frequencies at $T=0$ (Ref. \cite{combescot}
Right:
superconducting $T_c$ in the canonical ETh. At weak coupling $\lambda$, $T_c \approx 0.25 e^{-1/\lambda}$.
 At strong coupling, $T_c$ follows Allen-Dynes dependence $T_c \approx 0.18 \Omega_0 \sqrt{\lambda}$ (the dashed line in the figure).
}
 \label{fig:eli}
\end{figure}

In the right panel of Fig.\ref{fig:eli}  we plot $T_c (\lambda)$, obtained by solving the gap equation numerically, along with its asymptotic form at large $\lambda$. We see that strong coupling behavior $T_c \approx 0.1827\  \Omega_0\  \sqrt{\lambda} = 0.1827\  \alpha$ sets in at $\lambda \sim 5$, and $T_c$ exceeds $\Omega_0$ at even larger $\lambda \sim 30$.  The weak coupling behavior holds up to $\lambda \sim 0.5$, so the intermediate regime between the two limits is rather wide.
At $\lambda =1$, the actual $T_c$ is about a half of each of the two asymptotic forms.

\section{The validity of the Eliashberg theory in the strong coupling limit}
\label{sec:beyond}

We now discuss the self-consistency  of ETh at $\lambda \gg 1$.   We assume that both $\Omega_0$ and $\alpha$ are much smaller than $E_F$, but
the ratio $(\alpha/\Omega_0)^2 = \lambda$ can be arbitrary.

The ETh in the weak coupling regime is justified by the following four observations:
\begin{enumerate}
\item Pairing comes from fermionic states near the Fermi level,
 where one can linearize the fermionic dispersion near $k_F$.
\item The fermions are much faster excitations than
the phonons, and one can factorize the momentum integration in the expressions for the self-energy and the pairing vertex.
\item The corrections to the fermion-boson coupling $\alpha$ are small and can be ignored.
\item The corrections to phonon propagator $V (\Omega)$, can also be ignored.
\end{enumerate}

We need to reexamine these four conditions in the case of strong coupling.

\subsection{Linearization of the fermionic dispersion near the Fermi surface}

 At strong coupling, $T_c$ in the ETh is of order $\alpha$, hence the fermions, relevant to the pairing, also have energies of order $\alpha$.
   One can use the linearized dispersion for these fermions if
 \beq
 \alpha \ll E_F \ .
 \label{18}
 \eeq
(We assume that the Fermi energy, $E_F$, and the bandwidth are of the same order).  Eq. (\ref{18}) is satisfied in most DQMC studies and in general is not an obstacle for the applicability of the ETh at strong coupling because  the frequency dependence of the interaction makes the frequency sum in the formula for $T_c$ convergent, hence typical $\omega_m$ relevant to superconductivity are of order $T_c$.
 Then typical energy deviations from the Fermi surface are of order $v_F |k-k_F| \sim T_c \ll E_F$.

\subsection{Factorization of momentum integration}

This issue is not relevant for the canonical ETh, but is important for a more generic case when the
phonon propagator has momentum dependence.  This holds for pairing by acoustic phonons, but also for the case
 of pairing by optical phons, when one includes the renormalization of the bosonic propagator.
 We again use the fact that frequencies relevant to pairing are of order $\omega_m \sim T_c \sim \alpha$. At such frequencies, the fermionic $ {\tilde \Sigma}^* (\omega_m)$ from Eq. (\ref{8}) is  ${\tilde \Sigma}^* (\omega_m) \alpha^2/T_c \sim \alpha$  comparable to $\omega_m$, hence for estimates fermions can be treated as free quasiparticles.  The factorization of the momentum integration is then guaranteed by the smallness of the ratio $v_s/v_F \sim \Omega_0/E_F$ both at weak and strong coupling.

\subsection{Vertex corrections.}

The commonly cited result due to Migdal~\cite{Migdal} is that in 3D  the  corrections to the fermion-boson interaction $\alpha$ (often called the vertex correction) is
 \beq
  \frac{\delta \alpha }{\alpha } \sim  \lambda \frac{\Omega_0}{E_F}
  \label{20}
  \eeq
i.e., any vertex correction is the product of $\lambda$ and the ratio $\Omega_0/E_F$. The latter  appears in (\ref{20}) because in the processes
that give rise to vertex corrections, fermions are vibrating near a phonon frequency, far away from their mass shell.  Note in passing that there is no $\Omega_0/E_F$ factor in the self-energy diagram because there an intermediate fermion is near its own mass shell, and a phonon just provides a static interaction between mass-shell fermions.

At weak coupling, vertex corrections are small because both $\lambda$ and $\Omega_0/E_F$ are small.  At strong coupling, $\lambda$ is large, and the strength of vertex corrections depends on the interplay between $\lambda$ and $\Omega_0/E_F$. Because $\lambda = \alpha^2/\Omega^2_0$, the strength of vertex corrections is
\beq
  \frac{\delta \alpha }{\alpha }  \sim \frac{\alpha^2}{\Omega_0 E_F}
\label{21}
\eeq
At $\Omega_0 \to 0$, vertex corrections  diverge,  but because $\alpha \ll E_F$, this happens only at truly small $\Omega_0 < \alpha^2/E_F$, when Eliashberg $T_c$ is already close to its value at $\Omega_0=0$. One can also reach the strong coupling limit of ETh by taking $E_F \to \infty$ first and $\Omega_0 \to 0$ after,  while keeping  $\alpha$ finite.
   In this approach,  $\delta \alpha /\alpha $ remains small as  $\lambda \to \infty$.

The analysis of the vertex correction is actually not
so straightforward and requires some care, particularly in 2D.
At zero momentum transfer, corrections to the fermion-boson vertex $\delta \alpha$  are related by a Ward identity to the fermionic self-energy: $\delta \alpha/\alpha = d ~\Sigma (\omega)/d \omega = \lambda $.
   This vertex correction is only small at weak coupling, but not at $\lambda > 1$.  The argument that vertex corrections nevertheless can be neglected even at $\lambda >1$
   is due to the fact that typical momentum transfers in the processes leading to the self-energy and the renormalization of the pairing vertex, are of order $k_F$, hence one needs to know $\delta \alpha/\alpha$ for a finite momentum transfer of order $k_F$.
     For a generic momentum transfer ${\bf q}$,
  \beq
  \frac{\delta\alpha}\alpha = \lambda f\left(\frac{v_F |q|}{\Omega_0}\right),
  \label{19}
  \eeq
  where $~~f(0) =1$ and $f(x \gg 1) \sim 1/x$.
For $q \sim k_F$, the argument of $f(x)$ is   $ x= v_F |q|/\Omega_0 \sim E_F/\Omega_0 \gg 1$.
Substituting $f(x \gg 1) \sim 1/x$   into (\ref{19}), we reproduce Eq. (\ref{20}).

\begin{figure}
\includegraphics[width=.8\columnwidth]{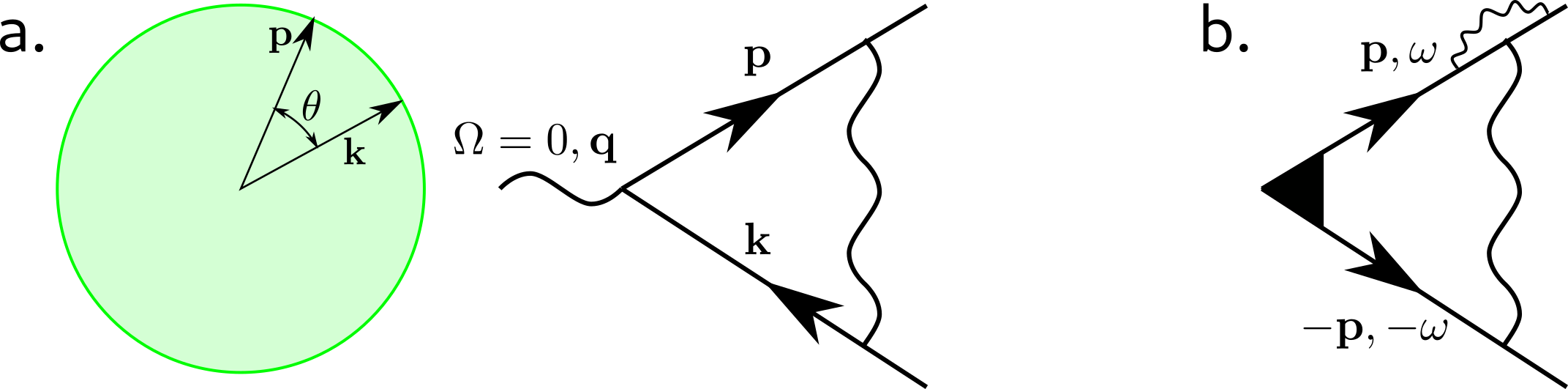}
\label{fig:vertex}
\caption{a) Lowest-order static vertex correction as a function of the momentum transfer ${\bf q} = {\bf k}-{\bf p}$ for particles on the Fermi surface (green circle). b) The diagram for the pairing vertex (black filled triangle) with the correction to the side vertex.  }
\end{figure}

In 2D, the situation is somewhat different. Evaluating the lowest-order vertex correction diagram, shown in Fig. \ref{fig:vertex}a,  at zero frequency transfer and small momentum transfer ${\bf q}= {\bf k} - {\bf p}$, and putting the external
 momenta ${\bf k}$ and ${\bf p}$  on the Fermi surface, such that $|q| = 2k_F \sin (\theta/2)$, where $\theta$ is the angle between ${\bf k}$ and ${\bf p}$, we obtain, at $|\sin (\theta/2)| > \Omega_0/E_F$,
 \beq
 \left|\frac{\delta \alpha }{\alpha }\right| = \lambda \frac{\Omega_0}{E_F}  \frac{\pi}{2 \sqrt{2} |\sin (\theta/2)|}
 \label{22}
 \eeq
 Substituting this into the pairing channel and comparing the renormalization of the pairing vertex with and without a vertex correction (Fig. \ref{fig:vertex}b) we find that adding a  vertex correction
   changes the renormalization of the pairing vertex by the factor $1+\tilde{Q}$, where
 \beq
 \tilde{Q} = \pi \sqrt{2} \lambda \frac{\Omega_0}{E_F} \log{\frac{E_F} {\Omega_0}}
\label{23}
\eeq
The $\tilde{Q}$ has the same factor $\lambda (\Omega_0/E_F)$ as the vertex correction in 3D,  but has an extra  logarithm.

\subsection{Renormalization of the bosonic propagator}
\label{sec:Rbosonic}

 The commonly used argument  to justify the neglect of the renormalization of the bosonic propagator in 3D is that the primary effect of such a renormalization is to add Landau damping to the phonon propagator.
 The Landau damping term  is the linear in $\Omega$ piece in the fermionic polarization bubble, which acts as a bosonic self-energy and converts $V (\Omega)$ into an effective
 \beq
 \frac{1}{V^{eff} (q, \Omega_m)} = \frac{1}{V (\Omega_m)} + \Pi (q, \Omega_m) \ .
 \label{24}
 \eeq
The Landau damping term $\Pi_L (q, \Omega_m )$ can be estimated by computing the particle-hole bubble:
\beq
\Pi_L (q,  \Omega) \sim \frac{|\Omega_m|}{v_F |q|} \ .
\label{25}
\eeq
Substituting into (\ref{24}) we obtain
\beq
V^{eff} (q,\Omega)  = \frac{\alpha^2}{\Omega^2_m + \Omega^2_0 + \frac{\alpha^2}{v_F |q|}  |\Omega_m|} \ .
\label{26}
\eeq
We now recall that at weak coupling the pairing is confined to frequencies  smaller than $\Omega_0$ and to momentum transfers of order $k_F$, while at strong coupling,  relevant  frequencies are of order $\alpha$ and relevant momenta are again of order $k_F$.
In both limits, the Landau damping term in the denominator in (\ref{26}) is parametrically smaller than max ($\Omega^2_m, \Omega^2_0$) and can be neglected. For $\lambda  = O(1)$,  typical
$|\Omega_m| \sim \Omega_0$ and  the Landau damping term is small by the same parameter $\Omega_0/E_F$, which makes vertex corrections small.
 In 2D, the effect of the Landau damping term has to be analyzed with extra care as the $1/|q|$ dependence in (\ref{26}) leads to an additional logarithm $\log {E_F/\Omega_0}$,
 as for the vertex corrections. Still,
 so long as the vertex corrections remain  parametrically small, the effect of the Landau damping term in $V^{eff} (q, \Omega)$ is also small.

This is, however, not the full story.  A simple inspection of the fermionic $\Pi (q,\Omega)$ shows that it also has the static contribution, $\Pi (q,0)$.
 The static polarization of free fermions in 2D  does not depend on $q$ up to $|q| = 2k_F$, i.e., for all momentum transfers relevant to pairing, and in our notations is equal to
 \beq
 \Pi (q,0) =  -2  \lambda \Omega^2_0
 \label{27}
\eeq
Substituting this $\Pi (q,0)$ into (\ref{24}) we obtain, even without the Landau damping,
\beq
V^{eff} (q, \Omega) = V^{eff} (\Omega) = \frac{\alpha^2}{\Omega^2_m + \Omega^2_0 (1-2 \lambda)}.
\label{28}
\eeq
We see  from (\ref{28}) that the renormalization of the bosonic propagator by the static polarization bubble
can only be neglected  for small $\lambda$.   Once $\lambda$ becomes of order one, this renormalization becomes crucial.
 Eq. (\ref{28}) shows that it {\it restricts the applicability of the canonical ETh to $\lambda <1/2$, which is well outside the strong coupling regime}.

Eq. (\ref{27}) was obtained by computing the polarization bubble for free fermions. For self-consistency, we need  to verify whether it remains valid for $\lambda \leq 1/2$. For this, we
  extend the calculation of the static polarization bubble  to higher orders by adding self-energy and vertex corrections inside the bubble.
 Self-energy corrections originate from inserting fermionic self-energy $\Sigma (\omega_m)$  into fermionic propagators in the bubble.  Using (\ref{28}) for the interaction, we obtain
 $\Sigma (\omega_m) = \lambda^{eff} \omega_m$, where
  \beq
  \lambda^{eff} = \frac{\lambda}{1-2 \lambda}.
 \label{31}
 \eeq
  The Green's function with $\Sigma (\omega)$ included is
 \beq
 G(k, \omega_m) = \frac{Z^{-1}}{i\omega_m - (v_F/Z) (k-k_F)}
 \label{29}
 \eeq
 where $Z = 1+\lambda^{eff}$. A
 calculation of the static particle-hole polarization bubble with these  $G(k,\omega)$ changes the free-fermion result for $\Pi (q,0)$ by
 a factor of $1/Z$.
 Vertex corrections inside the bubble in turn form a ladder series in $\lambda^{eff}/(1+ \lambda^{eff}) = (Z-1)/Z$ and change the free fermion result for $\Pi (q,0)$ by $\Gamma = 1/(1-(Z-1)/Z) = Z$.
   This result can be also obtained using the Ward identity $\Gamma = 1 + d \Sigma (\omega)/d \omega = 1+ \lambda^{eff} = Z$.
   Combining self-energy and vertex corrections we see that the factor $Z$  cancels out, i.e., $\Pi (q,0)$ remains the same as for free fermions. Thus, Eq. (\ref{28}) for  $V^{eff} (\Omega)$
    holds for $\lambda = O(1)$.
  Beyond ladder approximation, the dressed polarization bubble does acquire some momentum dependence.   In isotropic systems the static $\Pi (q,0)$ is generally peaked at $q=0$, in a lattice system it likely has a maximum at
   finite momenta.  In the last case, the vanishing of the mass term in $V^{eff} (q, \Omega)$ signals an instability towards CDW order with a particular ${\bf q}$.   In any case, the canonical Eliashberg theory becomes unstable at $\lambda = O(1)$.

 \section{Effective Eliashberg theory}
\label{sec:effective}

\subsection{Isotropic 2D systems}
\label{sec:isotr}
 Let us neglect for a moment  possible momentum dependence of $\Pi (q, 0)$  and use  Eq. (\ref{28}) for the phonon susceptibility. We see from (\ref{28})
 that the renormalization of the bosonic propagator can be absorbed into the  effective frequency $\Omega^{eff}_0 = \Omega_0 (1-2 \lambda)^{1/2}$. The new coupling $\lambda^{eff}$ is expressed via $\Omega^{eff}_0$ in the same way as without this renormalization, i.e., $\lambda^{eff} = \alpha^2/(\Omega^{eff}_0)^2$.  One can then introduce an {\it effective} ETh with $\Omega^{eff}_0$ instead of $\Omega_0$ and $\lambda^{eff}$ instead of $\lambda$. All expressions, which we earlier obtained for the canonical ETh are also valid for the effective ETh, but in the effective ETh the strong coupling regime does develop near $\lambda =1/2$.  In particular, Eliashberg $T_c \approx 0.1827 \alpha$.   For $\lambda \approx 1/2$, this $T_c$ is much larger than $\Omega^{eff}_0$ ($T_c = 0.1827 \Omega^{eff}_0 \sqrt{\lambda^{eff}}$).   At the same time, this $T_c$ can be equivalently re-expressed as  $T_c  \approx 0.13 \Omega_0$, i.e.,  it is only a fraction of the bare $\Omega_0$.
 Vertex corrections change the pairing interaction by $1 +\tilde{Q}^{eff}$, where
  \beq
\tilde{Q}^{eff} = \pi \sqrt{2} \lambda^{eff} \frac{\Omega^{eff}_0}{E_F} \log{\frac{E_F} {\Omega^{eff}_0}}
 \label{32}
 \eeq
For small $\Omega_0/E_F$, vertex corrections remain small for almost all $\lambda <1/2$, except for the immediate vicinity of $\lambda =1/2$, where the effective ETh breaks down.

We next include the momentum dependence of $\Pi (q,0)$.
 In an isotropic 2D system the momentum dependence comes from higher-order diagrams for the
   polarization bubble~\cite{Maslov_2017}, the same that give rise to the Kohn-Luttinger effect in 2D~\cite{Chubukov_1992}.
   We assume that $\Pi (q,0)$ has the smallest value at $q=0$. At the minimum,  $\Pi (0,0) \sim \lambda \Omega^2_0$, like in (\ref{27}), but with a different prefactor.
     Expanding around $q=0$ and using $|{\bf q}| = 2k_F \sin{\theta/2}$ for ${\bf q}$ between fermions on the Fermi surface, we obtain, neglecting the Landau damping,
    \beq
    V^{eff} (\Omega_m, \theta) = \frac{\alpha^2}{\Omega^2_m + (\Omega^{eff}_0)^2  + \beta^2 \sin^2{\theta/2}}.
     \label{nn_5}
     \eeq
  where
     $\Omega^{eff}_0 = \Omega_0 (1 - \lambda /\lambda_{cr})^{1/2}$ with
    $\lambda_{cr} = O(1)$, and $\beta$ sets the energy scale for the momentum dependence. Because the momentum dependence comes from fermions, $\beta$ is of order $E_F$, although
    the numerical prefactor is likely quite small in 2D (Ref. \cite{Maslov_2017}). In this respect, the ratio $\alpha/\beta$ can still be large even when $\alpha \ll E_F$.
   The self-energy  in the normal state at $T=0$ is
    \beq
    \Sigma (\omega_m)  = \frac{\alpha^2}{\Omega^{eff}_0} \int_0^{\omega_m/\Omega^{eff}_0} \frac{dx}{[(x^2+1)(x^2 +1 +{\bar \beta}^2)]^{1/2}}
    \label{nn_3_1}
    \eeq
     where ${\bar \beta}^2 = (\beta/\Omega^{eff}_0)^2$. At small $\omega_m$,  $\Sigma (\omega_m)  = \lambda^{eff} \omega_m$, where $\lambda^{eff} = (\alpha^2/(\Omega^{eff}_0))^2 /(1+ {\bar \beta}^2)^{1/2}$.  The same $\lambda^{eff}$ determines the self-energy at $T \neq 0$ at the first fermionic Matsubara frequency $\Sigma (\pi T) = \pi T \lambda^{eff}$.
 We plot $\Sigma (\omega_m)$ from (\ref{nn_3_1}) in Fig. \ref{fig:self_energy}.
  Comparing it with $\Sigma(\omega_m)$ for $\beta =0$ we see that the functional forms are similar, but the variation of $\Sigma(\omega_m)$ between small and large  $\omega_m/\Omega^{eff}_0$ gets smaller.

  The gap equation also get modified due to the different form of the self-energy and because the
    gap equation now contains an effective local interaction
\beq
V^{eff}_L (\Omega_m) = \langle V^{eff} (\Omega_m, \theta) \rangle
\label{33}
\eeq
 where the averaging is over the Fermi surface.  This effective interaction has a weaker dependence on frequency than  when $V^{eff} (\Omega_m)$ was independent of $q$.
  For $V^{eff} (\Omega_m, \theta)$ given by (\ref{nn_5}), $V^{eff}_L (\omega_m) = \alpha^2/((\Omega^2_m + (\Omega^{eff}_0)^2)(\Omega^2_m + (\Omega^{eff}_0)^2 + \beta^2))^{1/2}$.
    The analysis of the pairing with $\Sigma (\omega_m)$ from (\ref{nn_3}) and $V^{eff}_L (\Omega_m)$ from (\ref{33}) shows~\cite{Chubukov_2019_this_volume} that $T_c$ still
  saturates at a finite value when $\Omega^{eff}_0 \to 0$. When $\beta \ll \alpha$, $T_c$ changes little compared to the case $\beta =0$. In the opposite limit
    $\beta \gg \alpha$,   the angle variations in $V^{eff} (\Omega_m, \theta)$, relevant to pairing, are small
     and $T_c$ gets reduced.  To find $T_c$ in this case we need to go one step back and reconsider the Landau damping term $\Pi_L$ in (\ref{25}).
     Earlier we neglected this term because for $\beta =0$ typical angle variations along the Fermi surface are of order one, and for these variations $\Pi_L$ is small compared to $\Omega^2_m$ for $\Omega_m$ relevant to pairing.  At small angle variations,  $\Pi_L \sim \alpha^2 |\Omega_m|/(E_F |\theta|)$  is larger and may become relevant.   A simple analysis shows that there are two regimes of system behavior, depending on how large $\beta$ is.  For $\alpha \ll \beta \ll (\alpha^2 E_F)^{1/3}$, the Landau damping term is still irrelevant, $V^{eff} (\Omega_m, \theta)$ is given by (\ref{nn_5}), and
      $T_c \sim \alpha^2 \beta$.   For larger $\beta$, when $\alpha \ll  (\alpha^2 E_F)^{1/3} \ll \beta$, the Landau damping term is more relevant than the bare $\Omega^2_m$ term, and
       $V^{eff} (\Omega_m, \theta)$ is given by
     \beq
    V^{eff} (\Omega_m, \theta) = \frac{\alpha^2}{(\Omega^{eff}_0)^2  + \beta^2 \sin^2{\theta/2} + \alpha^2 \frac{|\Omega_m|}{2 k_F v_F |\sin{\theta/2}|}}.
     \label{nn_5_2}
     \eeq
    and $T_c$ is further reduced to  $T_c \sim (\alpha^2 /\beta) (\alpha^2 E_F/\beta^3)$.
     The effective interaction (\ref{nn_5_2}) has been analyzed in some detail in the context of purely electronic pairing by Ising-nematic fluctuations
      (see ~\cite{Chubukov_2019_this_volume} and references therein).

\subsection{2D lattice systems}

For fermions on a lattice  $\Pi (q,0)$  is generally peaked at some finite $q=q_0$.
   In this situation, $\Sigma (k_F, \omega_m)$ depends on the position of ${\bf k}_F$ on the Fermi surface. At weak coupling, the gap equation can be analyzed by restricting to the regions near "hot spots" -
   points on the Fermi surface separated by $q_0$.  At strong coupling, the whole Fermi surface becomes hot, and in general one
    cannot express the gap equation in  terms of local effective interaction, averaged over the Fermi surface. Instead, one has
    to solve the full integral gap equation in both momentum and frequency
   \cite{Metlitski2010,Wang2013,Abanov2008,Scalapino2012}.
   Alternatively, one can apply an approximate computation scheme: approximate the fermionic polarization $\Pi (q, \Omega_m)$ by a single bubble,  made out of dressed fermions
 and compute $\Pi (q, \Omega_m)$, the fermionic self-energy $\Sigma ({\bf k}_F,\omega_m)$, and  $V^{eff} (q, \Omega) = \alpha^2 (\Omega^2_m + \Omega^2_0 + \Pi (q, \Omega_m))^{-1}$ self-consistently.
    One then substitutes $V^{eff} (q,\omega_m)$ and $\Sigma ({\bf k}_F,\omega_m)$ into the gap equation, projects the pairing onto the $s-$wave channel, and obtains
   $T_c$ and $\Delta (\omega_m)$ below $T_c$.
 This is not a rigorous procedure because the self-consistent scheme neglects  higher-order vertex corrections to the polarization bubble, which are technically relevant for $\lambda = O(1)$, but it captures the key features of the evolution of $T_c$ near a point where $\Omega^{eff}_0$ softens at $q = q_0$. We call this computational scheme an extended ET.  It is quite similar to the fluctuation exchange approximation used to study spin-fluctuation mediated $d-$wave superconductivity (see, e.g., Ref. \cite{Manske_book}).

\begin{figure}[t!]
\begin{center}
\includegraphics[width=0.49\textwidth]{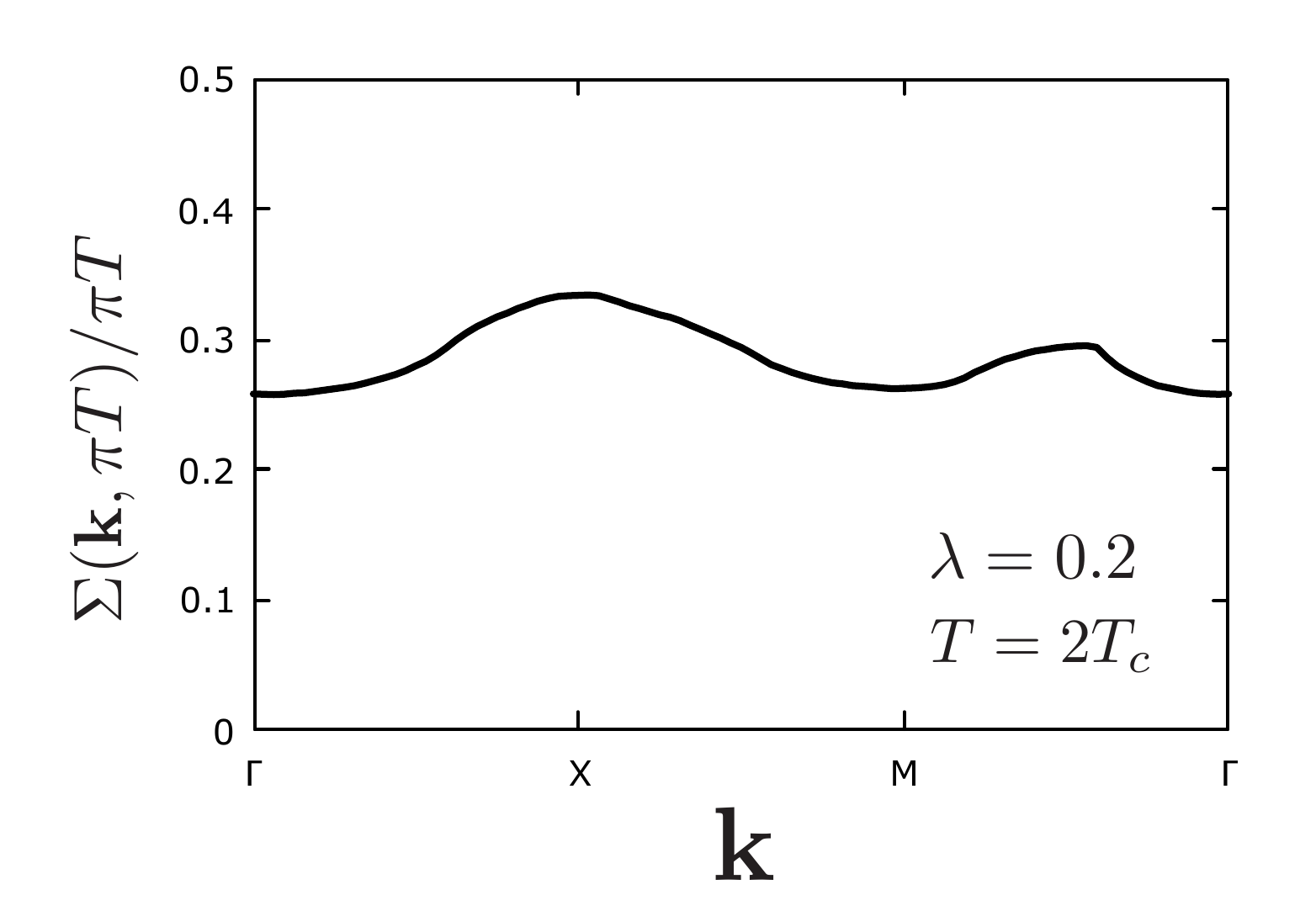}
\includegraphics[width=0.49\textwidth]{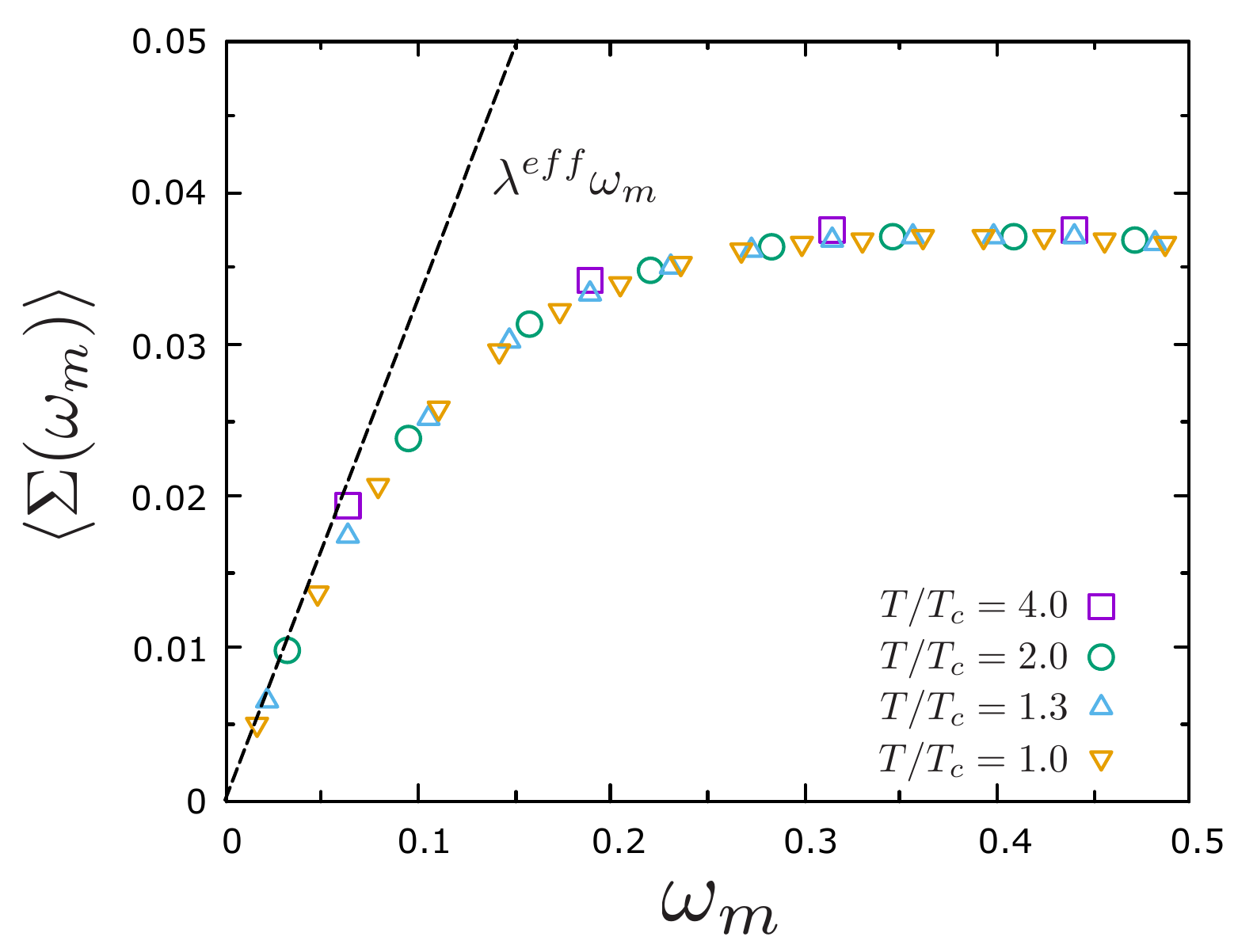}
\includegraphics[width=0.49\textwidth]{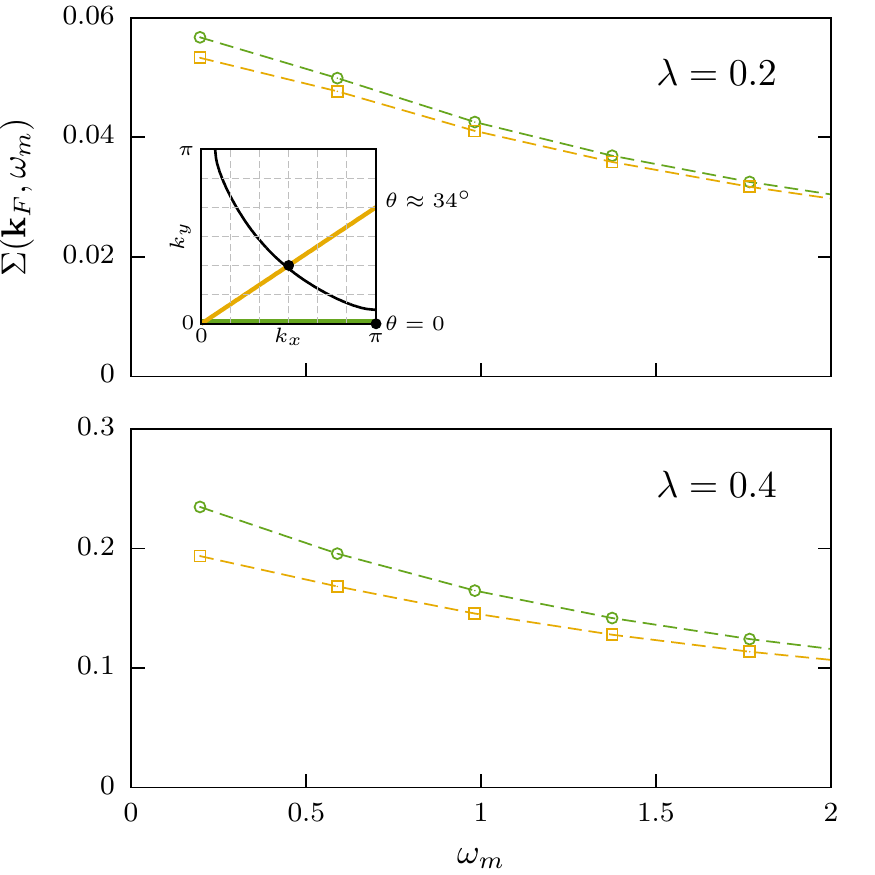}
\includegraphics[width=0.49\textwidth]{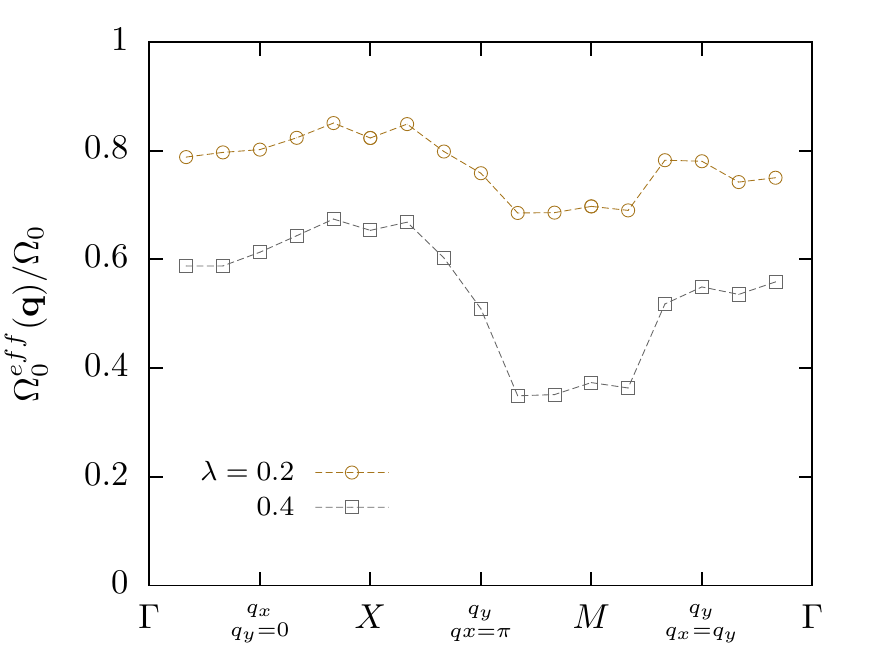}
\end{center}
\caption{Results for the effective ETh for  the tight-binding $t-t'$ model with nearest-neighbor hopping $t$ and next-nearest-neighbor $t' = -0.3t$. (a) The self-energy
$\Sigma({\bf k}, \pi T)$, normalized by $\pi T$, for the ${\bf k}$ path indicated on the horizontal axis through the Brillouin zone. (b) The self-energy, averaged over the Fermi surface, as a function of $\omega_m$.
 The averaged self-energy behaves as $\lambda^{eff} \omega_m$ at small frequencies and saturates at higher $\omega_m$.
 (c) The self-energy $\Sigma({\bf k}_F, \pi T)$ for two directions on the Fermi surface (shown in the insert) for two different values of $\lambda$.  (d) The ${\bf q}$-dependence of the effective $\Omega^{eff}_0 ({\bf q})/\Omega_0$ for two values of $\lambda$. Self-energy is units of the hopping $t$. For our choice of fermionic density, $E_F  \approx 1.7 t$. }
\label{fig:eliash_sig}
\end{figure}

\begin{figure}[t!]
\begin{center}
\includegraphics[width=0.5\textwidth]{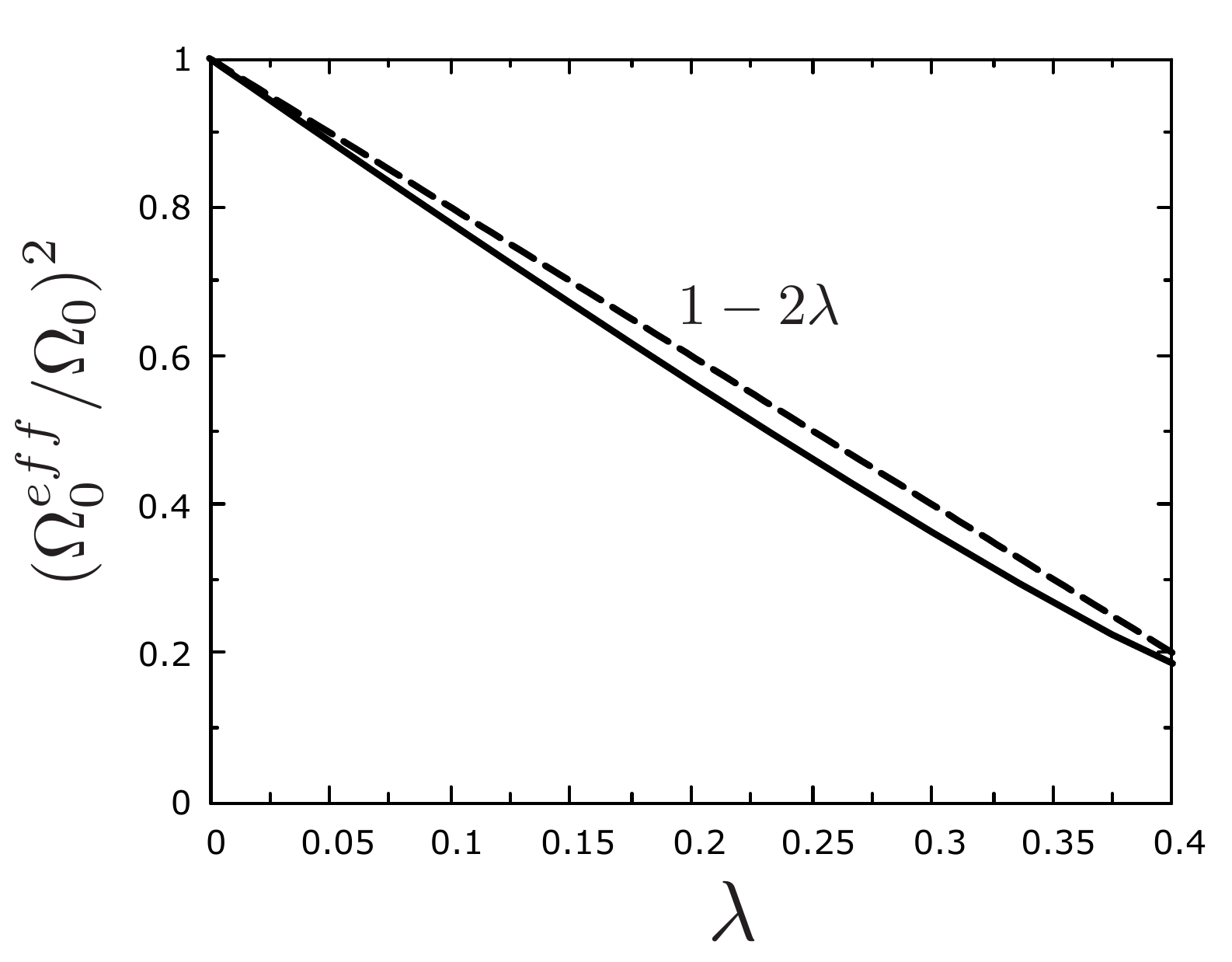}
\end{center}
\caption{The square of the ratio of "averaged" effective phonon frequency and the bare $\Omega_0$ (see text). For a momentum-independent interaction,
$(\Omega^{eff}_0/\Omega_0)^{2} = 1-2\lambda$, where $\lambda$ is the bare dimensionless fermion-boson coupling (dashed line in the Figure). The actual dependence (solid line) is almost the same.}
\label{fig:eliash_omega}
\end{figure}

\begin{figure}[t!]
\begin{center}
\includegraphics[width=\textwidth]{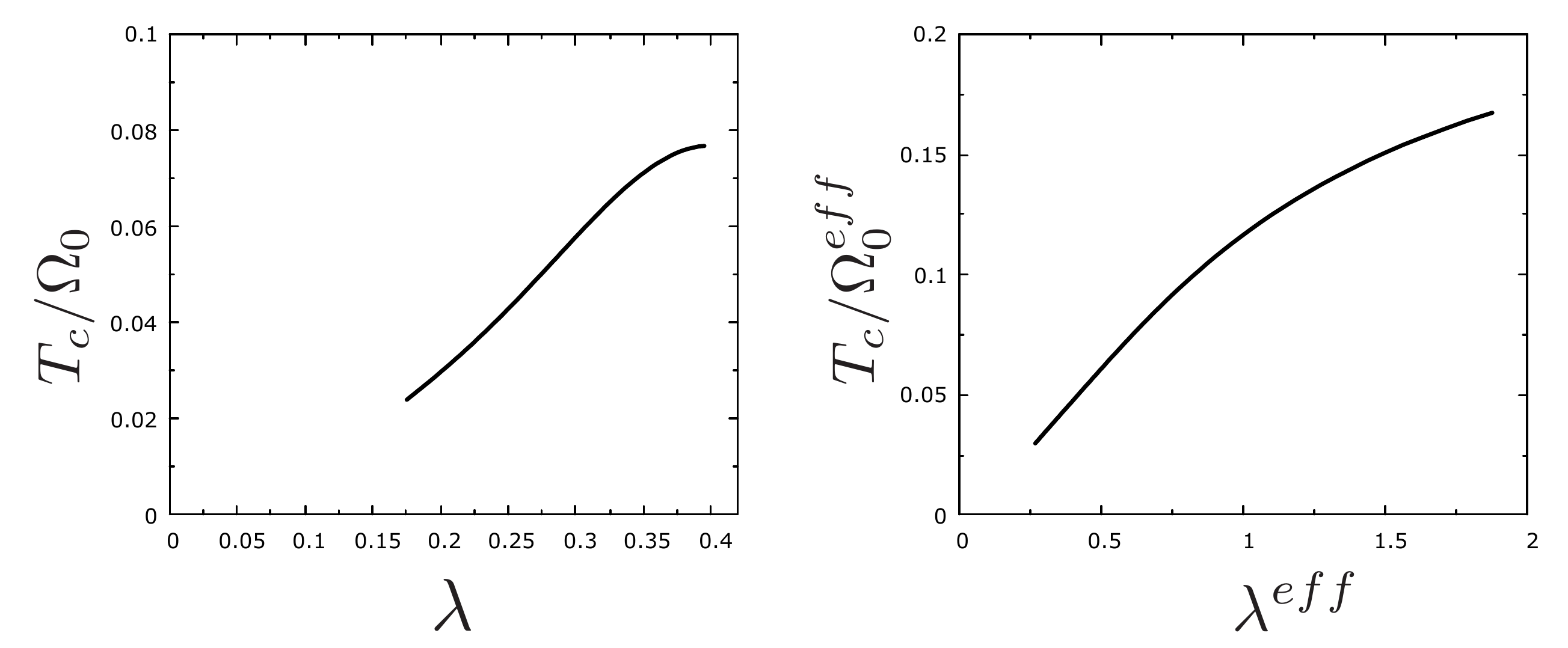}
\end{center}
\caption{Superconducting $T_c$ for the extended ETh and $t-t'$ dispersion. Left panel: $T_c/\Omega_0$ versus $\lambda$. Right panel: the same plot, but in terms of the effective parameters
$\Omega^{eff}_0$ and $\lambda^{eff}$. $T_c$ in the right panel roughly follows $\sqrt{\lambda^{eff}}$ behavior. Note that $T_c \ll \Omega_0$, and for realistic $\lambda_{eff}$ is remains smaller than $\Omega^{eff}_0$. }
\label{fig:eliash_tc}
\end{figure}

 We show the results obtained within the extended ETh in Figs.  \ref{fig:eliash_sig},
 \ref{fig:eliash_omega}, and \ref{fig:eliash_tc}. We consider a tight-binding model of fermions with  nearest-neighbor hopping $t$ and next-nearest-neighbor hopping $t'/t=-0.3$. We fix the electron density  $n=0.8$. This yields $E_F \approx 1.7t$.
 In Fig.  \ref{fig:eliash_sig}a we show $\Sigma({\bf k}, \pi T)$, plotted along a path in the Brillouin zone. In general, $\Sigma({\bf k},_F \pi T)$ determines the effective coupling $\lambda^{eff} ({\bf k}_F)$  via
   $\lambda^{eff} ({\bf k}_F)  = \Sigma ({\bf k}_F, \omega_m)/\omega_m$ at the smallest $\omega_m$.
   In a lattice system,   $\lambda^{eff} ({\bf k}_F)$ does in general depend on the location of ${\bf k}$ along the Fermi surface.  We see, however, that the full $\bf k$-dependence of $\Sigma$ is quite modest. In Fig.  \ref{fig:eliash_sig}b we show the frequency dependence of the self-energy, averaged over the Fermi surface. Frequencies $\omega_m$ are in units of the hopping $t = 0.6 E_F$.  Temperatures for this plot are much smaller than $t$, hence, to high accuracy,
 Matsubara frequency is a continuous variable, i.e., the self-energy is the same as at $T=0$.
  This is also evident from the fact that  the self-energy in  Fig.  \ref{fig:eliash_sig}b is very weakly $T$-dependent.
  In Fig.  \ref{fig:eliash_sig}b the dashed line has slope $\lambda^{eff}$, as defined by Eq. \ref{31}.
    Comparing
 $\langle\Sigma (\omega_m) \rangle$ with the one for the rotationally invariant case from Eq. (\ref{nn_3})
    (Fig. \ref{fig:self_energy}) we see that they are quite similar, just the overall variation of $\langle\Sigma (\omega_m) \rangle$ is a bit smaller for the same initial slope.
In the two other panels of this figure we show $\Sigma ({\bf k}_F, \omega_m)$ as a function of frequency for two directions on the Fermi surface, and the ${\bf q}$-dependence of the effective bosonic energy
$\Omega^{eff}_0 ({\bf q})$. The latter quantity is defined as
 $\Omega^{eff}_0({\bf q}) = \alpha/(V^{eff} (0, {\bf q}))^{1/2})$, where $V^{eff} (0, {\bf q})$ is the momentum-dependent static interaction.

 In Fig. \ref{fig:eliash_omega} we show the square of the ratio of the "averaged" effective bosonic energy $\Omega^{eff}_0$ and the bare $\Omega_0$:
 $(\Omega^{eff}_0/\Omega_0)^{2} = \alpha^2/(\Omega^2 _0 V^{eff}_L (0))$, where $V^{eff}_L (0)$ is the static interaction, integrated over the Fermi surface.
   If there was  no angle dependence of $V^{eff}$,  we would have  $(\Omega^{eff}_0/\Omega_0)^2  = 1 -2 \lambda$.  We see a very similar behavior within the self-consistent scheme, roughly up to $\lambda \sim 0.4$ (the best fit yields $2.13$ instead of $2$). At larger $\lambda$, the deviations start to grow.

 We show superconducting $T_c$ in Fig. \ref{fig:eliash_tc}. We see that $T_c$ increases with increasing $\lambda^{eff}$ and saturates at a finite value of order $\alpha$ when $\lambda^{eff}$ diverges.
 (Measured in units of the averaged $\Omega^{eff}_0$, $T_c$  does follow $\sqrt{\lambda^{eff}}$ behavior).
 This is quite similar to the behavior in Fig. \ref{fig:eli}. The numbers are also quite similar, when expressed in appropriate units: for e.g., $\lambda =2$, $T_c/\Omega_0$ in Fig. \ref{fig:eli} is about $0.2$, while for
  $\lambda^{eff} =2$, $T_c/\Omega^{eff}_0$ in Fig. \ref{fig:eliash_tc} is about $0.18$.

    Good agreement between the self-consistent calculation for the lattice model and the effective ETh with $\Omega^{eff}_0 = \Omega_0 (1-2\lambda)^{1/2}$ and $\lambda^{eff} = \lambda/(1-2 \lambda)$
    implies that, at least for the band structure used here, the effect of momentum dependence of the effective interaction is rather mild. To get an estimate, we approximated static $V^{eff} (0, {\bf q})$ by Eq. (\ref{nn_5}) and extracted $\beta/\alpha$ by fitting $\Omega^{eff} ({\bf q})$  in Fig. \ref{fig:eliash_sig}d. We found that $\alpha$ and $\beta$ are comparable: $\beta \sim 0.5 \Omega_0$ and $\alpha \sim 0.6 \Omega_0$.  In Sec. (\ref{sec:isotr}) we found that in this situation, $T_c$ is close to the  result for momentum-independent interaction, consistent with Fig. \ref{fig:eliash_tc}

 We  emphasize that although at $\lambda^{eff} =2$ the effective ETh approaches the strong coupling regime, $T_c$ is still much smaller than both the averaged $\Omega^{eff}_0$ and the variation of $\Omega^{eff}_0 ({\bf q})$ along the Fermi surface.  Like we said,  $T_c \sim 0.1827 \Omega^{eff}_0 \sqrt{\lambda^{eff}}$  exceeds $\Omega^{eff}_0$ only at $\lambda^{eff} >30$, which holds only extremely close to the point where $\Omega^{eff}_0$ vanishes.

\section{The validity of  Migdal-Eliashberg theory at $T > T_c$ }
\label{sec:normal}

We now briefly discuss the validity of a more general
 Migdal-Eliashberg theory
 for the electron-phonon interaction in the normal state $T > T_c$.
We argue that here the situation is more drastic because of thermal fluctuations. For the ETh of s-wave superconductivity, the contributions from thermal fluctuations to the fermionic self-energy and  the pairing vertex cancel because they effectively act as non-magnetic impurities.  However, for the normal state, the thermal self-energy plays a crucial role. The self-energy due to thermal fluctuations (the contribution from zero bosonic Matsubara frequency in (\ref{nn_2})) is computed differently from the self-energy at $T=0$  because the factorization of the momentum integration does not work for thermal fluctuations.  For  small enough $\Omega^{eff}_0$
  the bosonic propagator, integrated over both components of a 2D momentum,  is still singular, and
  to first approximation,
 \beq
 \Sigma_{th} (k, \omega) \sim T G(k, \omega) \lambda_T,
  \label{nnn_1}
  \eeq
  where $\lambda_T$ diverges at $\lambda = \lambda_{cr} (T)$, albeit more weakly than $\lambda^{eff}$. Such a self-energy, not included in the ETh, gives rise to precursors of the ordered state. The precursors  develop at $\lambda^* (T) < \lambda_{cr} (T)$  and shift the spectral weight
   from low-frequencies to a finite $|\omega| \sim (T \lambda_T)^{1/2}$. This changes the form of the spectral function and other observables and invalidates the ETh.
    The width of the precursor region increases with $T$.

  \begin{figure}[t!]
\includegraphics[width=0.5\textwidth]{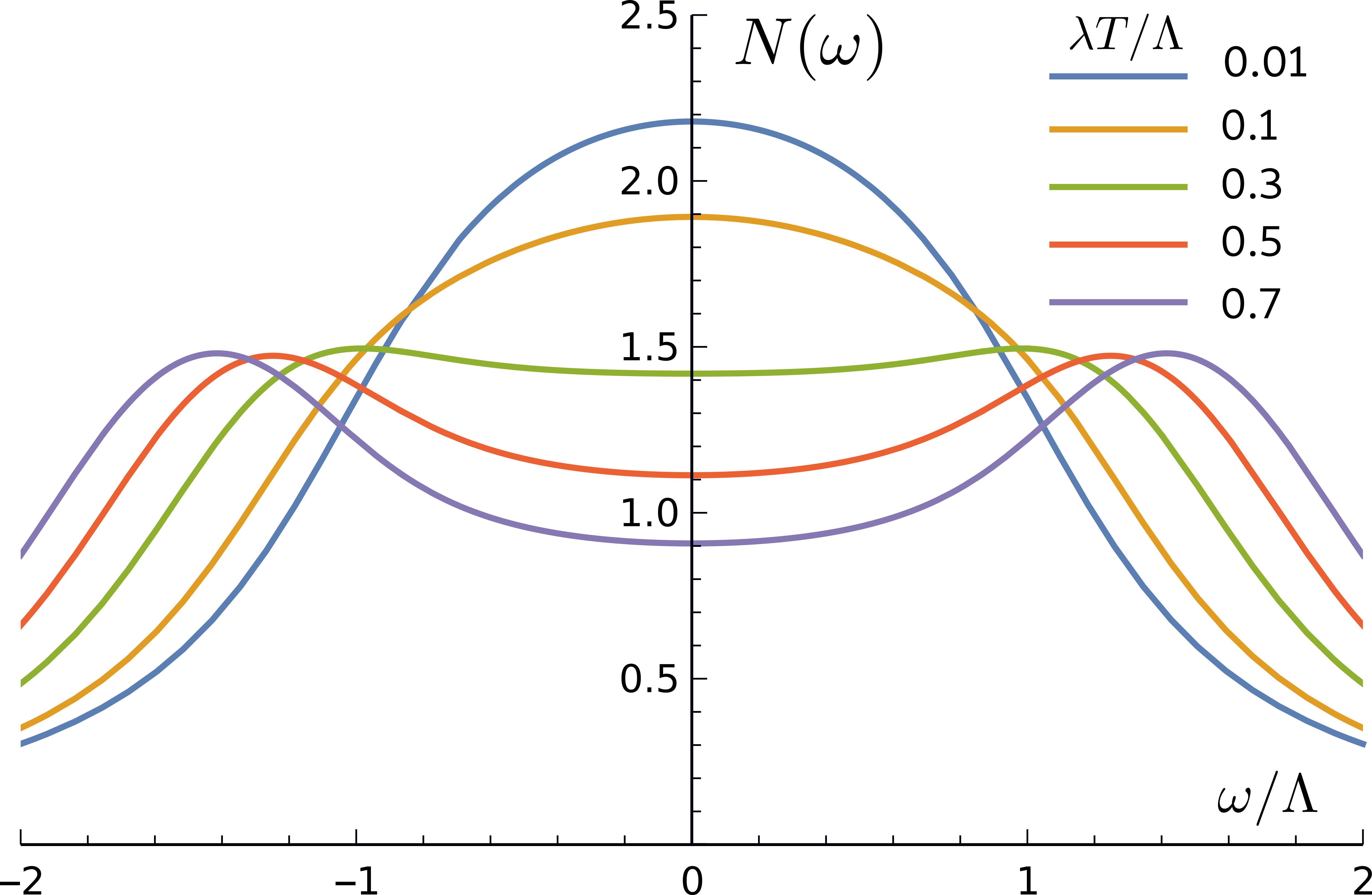}
\caption{Fermionic density of states $N(\omega)$ in the normal state, due to thermal fluctuations (Eq. (\ref{nnn_1_1}).
Frequency is in units of $\Lambda$, equal to a half of the bandwidth. The results are for $\lambda T /\Lambda = 0.01,0.1,0.3,0.5,0.7$. As $T$ increases, the maximum of $N(\omega)$ shifts to a finite frequency, and the system develops pseudogap behavior due to thermal fluctuations. This physics is outside Migdal-Eliashberg theory of the normal state.  We set the broadening $\delta =0.5 \Lambda$. }
\label{fig:thermal_dos}
\end{figure}
The effects of thermal fluctuations  can be analyzed more clearly if we choose another path to take the limit
$\Omega^{eff}_0 \to 0$,
as was done in the DQMC studies.  Previously  we kept the overall factor $\alpha^2$ in the bosonic propagator (\ref{aa_1})
 finite.  Then  $\lambda^{eff} = \alpha^2/(\Omega^{eff}_0)^2$ diverges when $\Omega^{eff}_0 \to 0$.  Let's now assume that  $\alpha^2$ by itself scales as $(\Omega^{eff}_0)^2$, such that $\alpha^2/(\Omega^{eff}_0)^2 = 1/{\bar k}$ remains finite.  The advantage of this approach is that at $\Omega^{eff}_0 \to 0$, the fermionic self-energy entirely comes from thermal fluctuations. Indeed, at finite $T$, the bosonic propagator at vanishing $\Omega^{eff}_0$,
 \beq
V^{eff} (\Omega_m) = \frac{1}{{\bar k}}  \frac{(\Omega^{eff}_0)^2}{4\pi^2 T^2 m^2 + (\Omega^{eff}_0)^2}
\label{eee_2}
 \eeq
 is finite only  for $m=0$.  There is no superconductivity, because the self-energy due to thermal fluctuations cancels out in the gap equations, but there are precursors to a charge-ordered state.

   Assume for simplicity that the non-interacting fermionic density of states is a constant in the frequency interval between $-\Lambda$ and $\Lambda$ and vanishes outside this interval.
    The one-loop retarded self-energy in real frequencies can be easily computed, and the result is
 \beq
 \Sigma (\omega) = - \frac{T}{{\bar k}} \log{\frac{\omega+ i \delta + \Lambda}{\omega+ i \delta - \Lambda}}
 \eeq
 At small $\omega$, $\Sigma (\omega) \approx i \pi T/{\bar k} - 2 T \omega/({\bar k}\Lambda)$. At large $\omega > \Lambda$, $\Sigma (\omega) \approx -2(T/{\bar k}) \Lambda/\omega$.
 The fermionic density of states is
 \beq
 N(\omega) = - {\text Im} Q (\omega), ~~ Q(\omega) =  \log{\frac{\omega + i \delta + \Lambda - (T/{\bar k}) \log{\frac{\omega+ i \delta + \Lambda}{\omega+ i \delta - \Lambda}}}{\omega + i \delta - \Lambda - (T/{\bar k}) \log{\frac{\omega+ i \delta + \Lambda}{\omega+ i \delta - \Lambda}}}}
 \label{nnn_1_1}
 \eeq
  In Fig. \ref{fig:thermal_dos} we plot $N(\omega)$ for several temperatures $T/({\bar k}\Lambda) = O(1)$.  We clearly see that $N(\omega)$ evolves as $T$ increases and at large enough $T$ develops precursors -- the peak in $N(\omega)$ shifts from $\omega=0$ to a finite frequency, of order $\Lambda$.   We emphasize that these precursors due to thermal fluctuations are beyond ETh.

  In the next Section we show that a similar behavior has been observed in DQMC studies. However, as will be explained further in the next Section, for $\lambda \gg 1$, the depression of spectral weight in the single-particle fermionic density of states is due to formation of localized bound pairs (bipolarons). The onset of a ``pseudogap" due to formation of pairs is more complex phenomenon than the one-loop effect
   that we discussed above.
 The main point of this Section, therefore, is just to illustrate how thermal fluctuations can invalidate the Migdal-Eliashberg theory, \textit{even for} $\lambda \lesssim 1$. We note in passing  that the effects of thermal fluctuations can be studied beyond one-loop order using a computational procedure  similar to the eikonal approximation in the scattering theory (see e.g., Ref. \cite{Ye2019} and references therein).

\section{Comparison with Monte-Carlo analysis }
\label{sec:QMC}

\subsection{Self-energy, bosonic propagator, and pairing susceptibility}

In this Section we compare the results obtained using the extended ETh with the results  of extensive Monte Carlo calculations for the Holstein model~ \cite{DQMC,QMC}. The model describes  tightly bound electrons on a 2D square lattice coupled to an optical phonon mode with frequency $\Omega_0$. The explicit form of the Hamiltonian is 	\beq
	H = \sum_{ij}t_{ij}c^\dag_{i\sigma}c_{j\sigma}  + \frac 12 \sum_i (\chi_0 p_i^2 + \chi_0^{-1}\Omega_0^2 x_i^2) + g \sum_{i\sigma} x_i c^\dag_{i\sigma} c_{i\sigma},
\label{qq_1}	
\eeq
where $c^\dag_{i\sigma}$ creates an electron at site $i$ with spin $\sigma$ and $x_i$ is the local oscillator displacement at site $i$ and $p_i$ is the conjugate momentum, $[x_i,p_j]=i\delta_{ij}$. We choose $t_{ij}$ with  nearest-neighbor hopping $t$ and next-nearest-neighbor hopping $t'/t=-0.3$. We fix the electron density  at $n=0.8$, in which case  $E_F \approx 1.7 t$. We present results for $\Omega_0/E_F = 0.1$.

In the notations of Eq. (\ref{qq_1}), the effective fermion-boson coupling $\alpha^2$ is expressed as
\beq
\alpha^2 = g^2 N_F \chi_0,
\label{qq_2}
\eeq
and the dimensionless coupling $\lambda$ is
\beq
\lambda = \frac{g^2 N_F \chi_0}{\Omega^2_0}
\label{qq_3}
\eeq

The focus in Ref.~\cite{DQMC} was on the breakdown of the ETh when the
 bare coupling $\lambda$  reaches some value $\lambda_{cr}$ of order one.
  In Ref.~\cite{DQMC} is was found that $\lambda_{cr}\approx 0.4$.
    DQMC analysis includes
  vertex corrections, hence $\lambda_{cr}$ in DQMC should be somewhat smaller than the one at which
    extended ETh breaks down.
   For $\lambda > \lambda_{cr}$,  DQMC study has found that  at finite $T$ electronic states are affected across the entire band and the low-energy spectrum changes dramatically from dressed electronic quasiparticles to bipolarons, which acquire a large effective mass and behave effectively as a classical lattice gas. Rather than superconducting, the bipolarons tend to form various commensurate charge-ordered states, or else phase separate.

Our focus here is superconductivity and we will first consider
  $\lambda < \lambda_{cr}$, where the ETh remains
  viable.  We will show that, in this regime, certain predictions of the extended ETh are in fact remarkably consistent with DQMC.

\begin{figure}[t!]
\begin{center}
\includegraphics[width=0.49\textwidth]{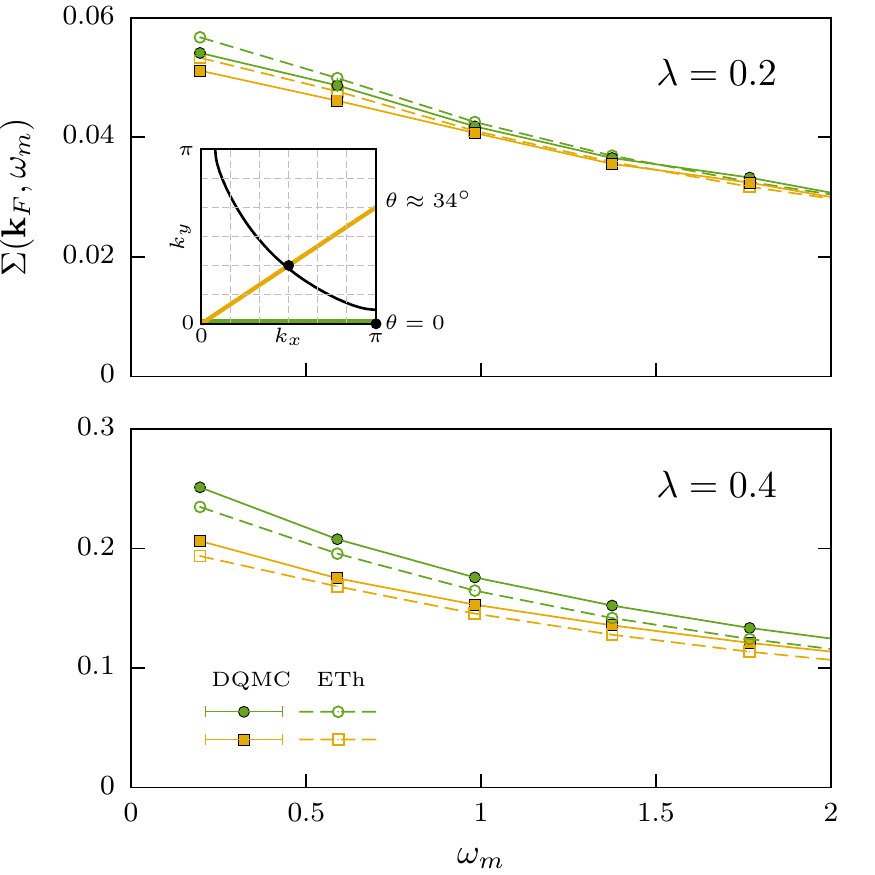}
\includegraphics[width=0.49\textwidth]{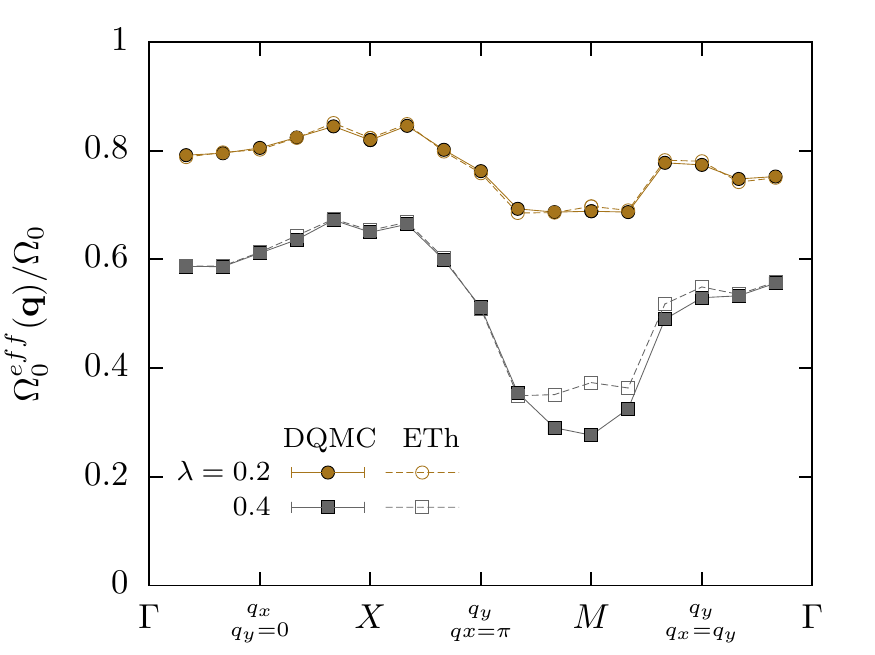}
\end{center}
\caption{Comparison of the results obtained within the extended ETh (empty circles/squares) and DQMC (filled circles/squares).
 Left panel: the self-energy. Right panel: the
ratio of the effective and the bare  phonon frequency, $\Omega^{eff}_0/\Omega_0$. In both figures the temperature is $T \approx E_F/25$.}
\label{fig:MC_comparisons}
\end{figure}

\begin{figure}[t!]
\begin{center}
\includegraphics[width=0.49\textwidth]{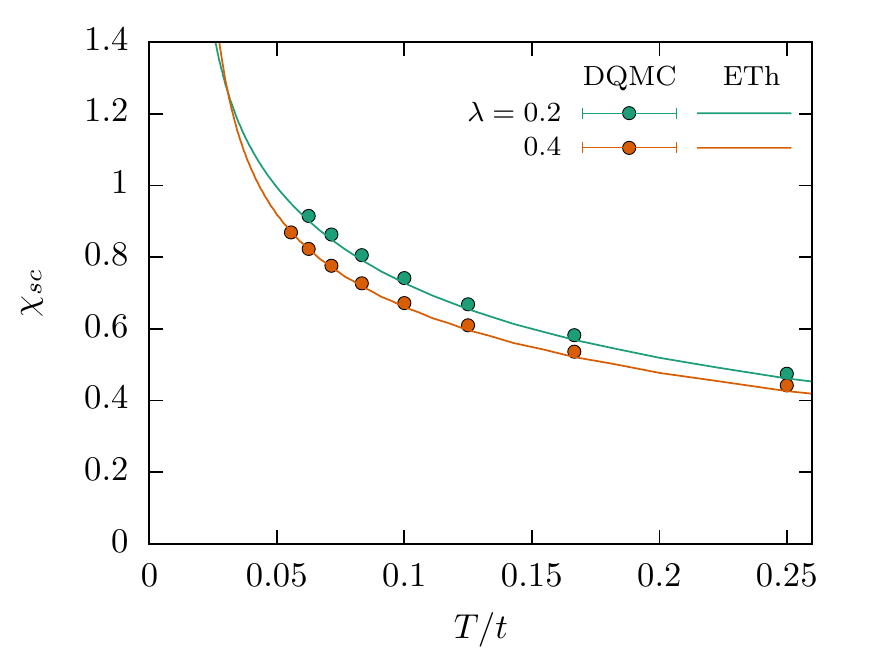}
\end{center}
\caption{Comparison of the results within the extended ETh (lines) and DQMC (dots)  for the
 static  $s$-wave pair susceptibility $\chi_{sc}$. }
\label{fig:MC_comparison_chisc}
\end{figure}

The normal state self-energy and the effective, $q-$dependent phonon frequency $\Omega^{eff}_0 (q)$ are shown in Fig.~\ref{fig:MC_comparisons} for temperature $T \approx E_F/25$, which is  the lowest temperature we were able to access by DQMC. Both are remarkably close to the ones obtained within the extended ETh (same as in Fig. \ref{fig:eliash_sig}d), which we also present in these figures.  Notice that the momentum dispersion is rather small for $\lambda =0.2$, but increases for $\lambda =0.4$.
For $\lambda = 0.4$, there is a noticeable difference between DQMC and extended ETh in a narrow range of q around $(\pi,\pi)$, this reflects an emerging problem in treating the tendency towards CDW
(Ref. \cite{QMC}.) In Fig.~\ref{fig:MC_comparison_chisc} we show the $s$-wave pair susceptibility $\chi_{sc}$, defined as
	\beq
	\chi_{sc} = \int_0^\beta d\tau ~ \langle \Delta(\tau)\Delta^\dag(0)\rangle, \quad \Delta^\dag = \frac 1L \sum_i c_{i\uparrow}^\dag c_{i\downarrow}^\dag,
	\eeq
and $L$ is the linear system size. The lines show $\chi_{sc}$, obtained within the
 extended ETh. We see that the extended ETh and DQMC yield almost identical results for $\chi_{sc}$ over
 the entire accessible temperature range.

\subsection{The full phase diagram of the Holstein model}

\begin{figure}
\includegraphics[width=0.49\textwidth]{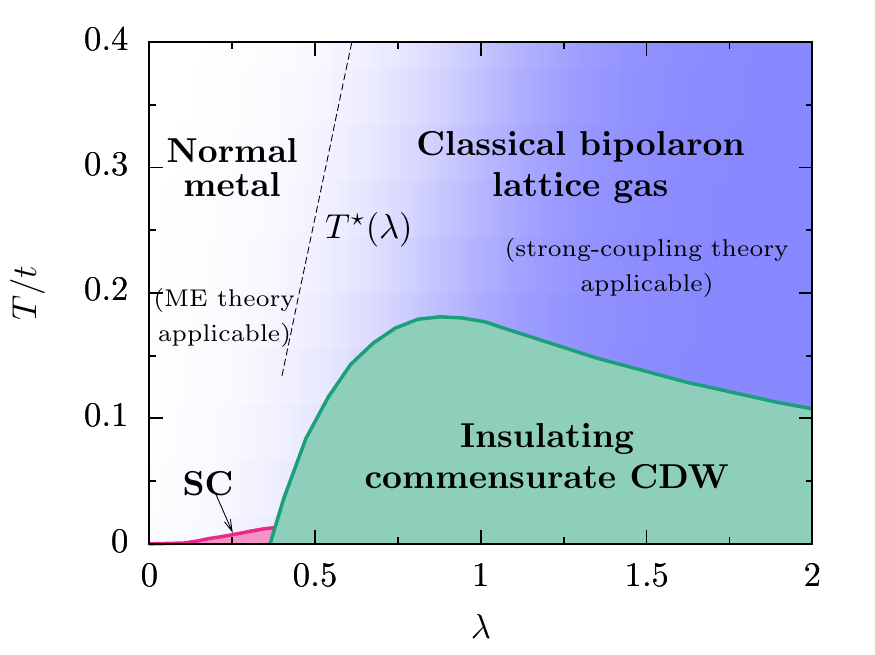}
\includegraphics[width=0.49\textwidth]{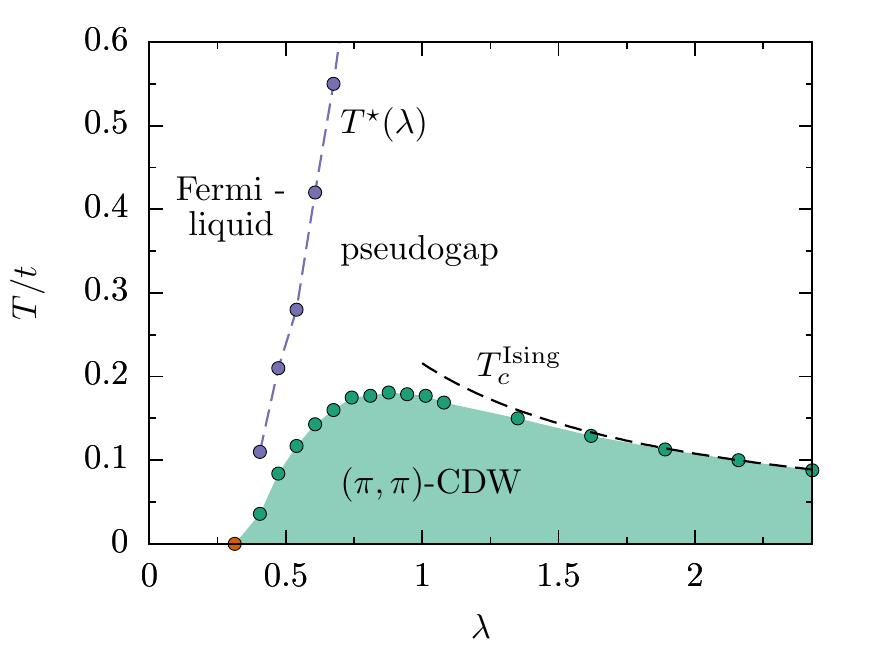}
\caption{The phase diagram emerging from DQMC studies.  Left panel: full DQMC calculations. At higher $T$, there is a wide crossover region around $T^* (\lambda)$,
 separating normal metal behavior, for which ETh (more accurately, Migdal-Eliashberg theory)  is applicable, and classical bipolaron lattice gas, for which Migdal-Eliashberg description is not applicable.  At low $T$ the system develops superconductivity for $\lambda \leq 0.4$ and commensurate CDW state at larger $\lambda$.
 Right panel -- the results of a separate DQMC study, in which the limit $\Omega^{eff}_0 \to 0$ has been taken such that the phonon stiffness, $k = \chi^{-1}_0(\Omega^{eff}_0)^2$, was kept fixed.
    In this particular limit  the dimensionless coupling  $\lambda = g^2 N_F \chi_0/(\Omega^{eff}_0)^2 =
   g^2 N_F/k$
   remains finite. There is no superconductivity in this case, but the CDW phase and the $T^* (\lambda)$ line are present.
     To the left of this line the system behaves as a Fermi liquid and Migdal-Eliashberg is applicable down to zero temperature. To the right, the single-particle spectral function develops a pseudogap. In this regime Migdal-Eliashberg theory becomes entirely inapplicable.}
\label{fig:global_phase_diag}
\end{figure}

In this Section we describe the global phase diagram of the Holstein model at $T > T_c$,  as a function of $\lambda$ and temperature $T$, in the limit $\Omega_0/E_F \ll 1$ (Ref. \cite{QMC}).  The schematic phase diagram in Fig.~\ref{fig:global_phase_diag} presents the summary of the results. The key finding, relevant to the current discussion,  is the existence of a crossover line $T^\star(\lambda)$, separating the phase diagram into two qualitatively distinct regions. To the left of the $T^\star$ line the ETh is both qualitatively and quantitatively accurate; to the right the ETh breaks down \textit{qualitatively}.
  In this last region the low-energy degrees of freedom at higher $T$ are bipolarons with a binding energy $\sim g^2 \chi_0/\Omega_0^2$ and there is a pseudogap to single-particle excitations. At lower $T$ the system has a tendency to form commensurate charge-ordered states, with a wave-vector unrelated to nesting vectors of the Fermi surface.

The schematic phase diagram of Fig.~\ref{fig:global_phase_diag} is based on the DQMC studies, described in the previous section, as well as a separate DQMC study, in which the limit $\Omega^{eff}_0 \to 0$ has been taken such that the phonon stiffness, $k = \chi^{-1}_0(\Omega^{eff}_0)^2$, was kept fixed.
    In this particular limit  the dimensionless coupling  $\lambda = g^2 N_F \chi_0/(\Omega^{eff}_0)^2 =
   g^2 N_F/k$
   remains finite. We modeled this approach in Sec. (\ref{sec:normal}).  The
  bosonic propagator is given by Eq. (\ref{eee_2}) with $k= {\bar k}/(g^2 N_F)$ and
  is non-vanishing
only at $\Omega_m =0$, i.e., only
  static, thermal fluctuations of the phonons contribute to the fermionic self-energy.
   The reason for working in this particular limit is that standard DQMC becomes computationally intractable as the coupling strength is increased. The simplification described here ameliorates those difficulties and gives access to the entire phase diagram. Moreover, the physics of the strong-coupling regime is expected to be largely insensitive to $\Omega_0$, so long as $\Omega_0 \ll E_F$. In the weak-coupling regime this limit should be quantitatively accurate in the regime $\Omega_0 \ll T \ll E_F$. This has also been verified by comparing with the full DQMC calculations with $\Omega_0/E_F = 0.1$, described in the previous section.
   Superconductivity is absent in the limit $\Omega_0 = 0$ because then $\Omega^{eff}_0$ also vanishes, and $V^{eff} (\Omega_m)$ has only the contribution from
     thermal fluctuations, which cancel out in the gap equation. For
        $\Omega_0/E_F=0.1$, $T_c$ is non-zero, but too low to be detected by DQMC.  However, given the quantitative reliability of ETh to the left of the $T^\star$ line (see in particular Fig.~\ref{fig:MC_comparison_chisc}), we can use it to reliably extrapolate to lower temperature and obtain estimates of $T_c$. This is the procedure by which the superconducting region of the phase diagram in Fig.\ref{fig:global_phase_diag} was obtained.

The results of such DQMC calculation for the case $\Omega_0 \to 0$ are shown in Fig.~\ref{fig:global_phase_diag}.
 The electronic band structure is the same as in the previous section.
 To the left of the $T^*$ line the system behaves as a Fermi liquid and is metallic down to zero temperature. To the right of the $T^\star$ line the single-particle spectral function develops a pseudogap. In this regime the ETh becomes entirely inapplicable.
  This is fully consistent with our analysis in Sec. \ref{sec:normal}.
  Remember that thermal fluctuations are not included into either canonical or effective ETh, so when these fluctuations becomes strong, ETh necessarily breaks down.

 At sufficiently low temperature below $T^\star$ there is a transition to a commensurate $(\pi,\pi)$ CDW state. The $T=0$ transition is first order, while all the observed finite temperature transitions appear to be continuous (presumably, the first order transition persists to some low but nonzero temperature). As explained in \cite{QMC}, to leading order in the strong-coupling expansion in powers of $1/\lambda$ the Holstein Hamiltonian in the limit $\Omega_0/E_F \ll1 $ maps to the antiferromagnetic Ising model in an external field. From this perspective, the $(\pi,\pi)$ transition is natural, corresponding to the commensurate, antiferromagnetic ordering transition of the Ising model at a temperature $T_c^\mathrm{Ising}$. Fig. \ref{fig:global_phase_diag} shows that $T_c^\mathrm{Ising}$, computed with parameters from the strong-coupling expansion, coincides accurately with the CDW transition temperature of the full Holstein model for $\lambda \gtrsim 1$.

\begin{figure}
\includegraphics{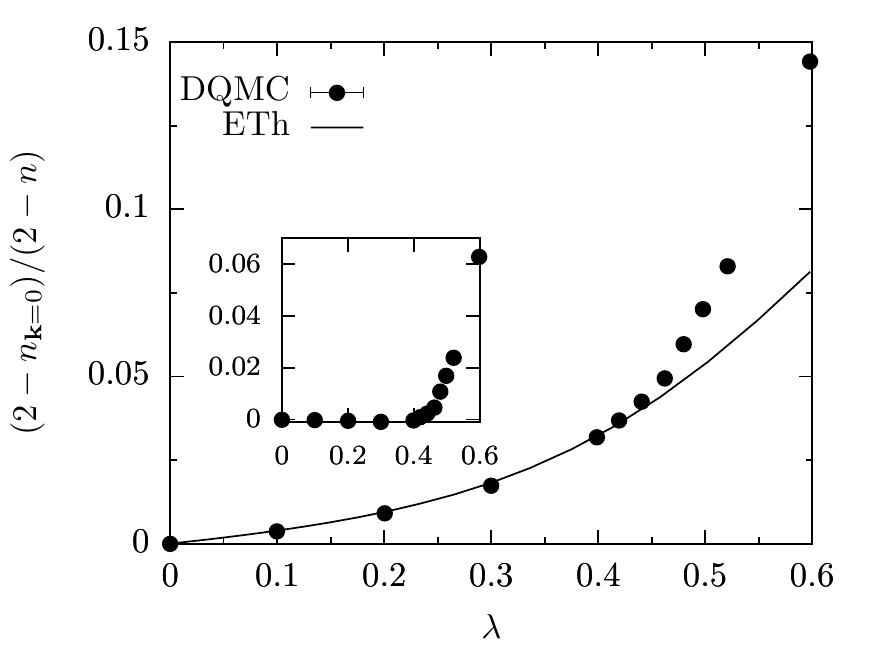}
\caption{The occupation number of the single-particle state at the bottom of the electron band, $n_{\mathbf k=0}$. Solid line -- the result within the extended ET, dots are DQMC results.  We see that DQMC and ETh results almost coincide for $\lambda <0.4$, but rapidly deviate for $\lambda >0.4$. Inset shows the different between DQMC and ET. Note the precipitous increase in the error for $\lambda \gtrsim 0.4$.}
\label{fig:nk0}
\end{figure}

To better understand the finite-temperature breakdown of ETh, we show in Fig.~\ref{fig:nk0} the occupation number of the single-particle state at the bottom of the electron band, $n_{\mathbf k=0}$. As already explained, in ETh one takes the bandwidth to infinity at the outset, focusing only on a narrow band of energy $\sim \Omega_0$ around $E_F$. This approximation becomes invalid when $g^2\chi_0/\Omega_0^2 \sim E_F$; i.e., when $\lambda = \mathcal O(1)$ ($N_F \sim 1/E_F$), at which point the entire electronic spectrum is rearranged. This effect is evident in Fig.~\ref{fig:nk0}, where we observe a precipitous change in the occupation of the electronic state deepest in the band.

\section{Summary}
\label{sec:conclusions}

In this work we  analyzed of validity of ETh of phonon-mediated superconductivity in 2D systems in light of recent extensive Monte-Carlo studies of the Holstein model.
For analytical analysis, we considered a model of fermions, coupled to a single Einstein phonon with frequency $\Omega_0$. The dimensionless coupling in this model is $\lambda = \alpha^2/\Omega^2_0$, where $\alpha$ (with dimension of energy) is the effective electron-phonon coupling, which incorporates fermionic density of states.

We found that:

\begin{enumerate}
\item The canonical ETh breaks down when the bare coupling reaches a critical value $\lambda_{cr} = O(1)$.
 At this value, the  would be Fermi liquid ground state in the absence of SC becomes unstable.
 To a good approximation, $\lambda_{cr} = 1/2$.

\item Near the instability, the phonon frequency softens, and the system enters
a strong coupling regime,
although the bare coupling is of order one.  In general, in this regime the dressed phonon propagator
     becomes momentum dependent and softens first either at $q=0$ (in a spatially isotropic system) or at a finite $q$ in a lattice system.  Away from the immediate vicinity of $\lambda_{cr}$, the $T=0$ properties of a would be normal state are
     approximately described by  an effective ETh with
      $\Omega^{eff}_0 = \Omega_0 (1-2\lambda)^{1/2}$ and $\lambda^{eff} = \lambda/(1-2\lambda)$.

\item
Superconductivity near the critical point can
plausibly be well described  within the strong coupling limit of the effective ETh.
 A characteristic temperature $T_c$, which may be better interpreted as an onset of pairing than the actual
  transition temperature,
   saturates to a finite value as the effective coupling diverges.
  For the isotropic dispersion, $T_c \approx 0.18 \alpha \approx 0.08 \Omega_0$. In a lattice system, the prefactor is generally a bit smaller. This $T_c$  is much smaller than $\Omega_0$ and is even smaller than $\Omega^{eff}_0$, except in the immediate vicinity of $\lambda_{cr}$.
  \item
  Effective ETh  breaks down at some $\lambda^* <\lambda_{cr}$, because  vertex corrections become large. In 2D vertex corrections are logarithmically enhanced compared to 3D case and are of order
  $(\alpha^2/(\Omega^{eff}_0 E_F) \log(E_F/\Omega^{eff}_0)$.  Still, for large $E_F$, ETh breaks only near $\lambda_{cr}$.
 \end{enumerate}

We emphasize that in our consideration we assumed that  at  $\lambda =\lambda_{cr}$ the system undergoes a conventional second-order transition, in which it becomes unstable towards
 a charge order, bilinear in fermions.  Such an order is accompanied by the softening  of a phonon mode at some $q=q_0$.  If, however, the $T=0$ transition is either first order, or is more complex (e.g., a multi-phonon propagator softens before a single-phonon one), the effective ETh breaks down
 at $\lambda^* < \lambda_{cr}$, even if  vertex corrections are still small at $\lambda^*$.   Also,  we assumed that  the electron-phonon coupling $\alpha$ is small compared to Fermi energy.  When $\alpha$ becomes comparable to $E_F$,  the effects associated with electron localization (Mott physics) becomes progressively more relevant. In this situation,  the region of applicability of both the canonical and the effective ETh shrinks, and for large enough $\alpha$ ETh becomes unapplicable.

\section{Discussion}
\label{sec:conclusions_1}

We view the present discussion as a step toward reconciling various different approaches to the problem of boson mediated superconductivity, but there are still aspects of the problem that look different when approached from different perspectives, and these need to be reconciled.  This will require further work.
 We now
  step back a bit to discuss the problem from a more general perspective to emphasize what we think are still vexed issues.

The Migdal approximation  involves neglecting all vertex corrections, which leads to a closed set of integral equations for the electron and phonon self energies, $\Sigma(\vec k,\omega)$ and $\Pi(\vec k,\omega)$.  If we  introduce Nambu spinors and allow for an anomalous term in the electron self-energy, the same set of integral relations give the Migdal-Eliashberg approximation for the properties of the superconducting state. There is a widely held belief that this approximation is valid for computing general features of the electron-phonon problem even if the dimensionless electron-phonon coupling, $\lambda$, is large so long as the ``Migdal parameter,'' $\lambda (\Omega_0/E_F)$, is sufficiently small.  Comparison between various quantities computed in the Migdal approximation and those computed by DQMC prove that this belief is wrong, and in the above we have identified analytically some of the ways in which this breakdown occurs for various ``normal state'' properties. It is important to stress that this breakdown
 occurs at temperatures  high enough that neither superconducting nor charge-density wave correlations extend over any significant range of distances, so it cannot be associated with the onset of an instability toward any of the relevant ordered ground-states - rather it is associated with the local physics of classical bipolaron formation.

However,  it is possible that - despite the fact that aspects of the electron self-energy (and many other features of the problem) are overall ill-accounted for by the diagrams that are summed in the Migdal-Eliashberg treatment, one might still be able to obtain reliable results from the same set of equations
for other properties, in particular  the superconducting $T_c$ and the superconducting gap structure below $T_c$.  While a priori this proposition sounds strange, the above analysis suggests that much that is missed in Migdal-Eliashberg approach is inessential for these specific features of the superconducting state.  To make this proposition more plausible, we remind the reader of a related case in which controlled calculations are possible, and where similar underlying mathematical structures account for this nonintuitive state of affairs.

Consider the case of electrons in high dimension $d > 2$ in the presence of a weak attractive interaction, $U$, and weak disorder:
\begin{itemize}
\item{\bf  Ignoring the effect of disorder,} the attractive interaction leads to the existence of electron-electron scattering which leads to a normal-state quasi-particle scattering rate, $1/\tau_{el-el} \sim U^2 T^2E_F^{-3}$, and a mean-field superconducting transition temperature that depends exponentially on $E_F/U$ as $\ln[T_{c0}/E_F]\sim - U/E_F$. Correspondingly, there is an exponentially small gap function that is approximately $\vec k$ and $\omega$ independent of magnitude $\Delta_0 \approx 3.53 T_{c0}\ll U < E_F$, and correspondingly an exponentially long superconducting coherence length, $\xi_0 =
     v_F/\Delta_0$. Moreover, the mean-field value of $T_c$ is accurate to exponential accuracy, as the Ginzburg parameter (which controls the range of $T$ in which fluctuations about the mean-field solution are significant) is itself exponentially small, $g = [\rho(E_F) \Delta_0 \xi_0^d]^{-1} \sim [k_F\xi_0]^{-(d-1)}$.

\item{\bf Ignoring the interactions}, we have a dirty metal with a quasiparticle scattering rate $1/\tau_{dis} \sim v_F/\ell$ where $\ell$ is the elastic mean-free path.  Naturally as the system is non-interacting, there can be no finite $T$ transitions, and since by assumption we are in $d>2$, the system remains metallic even as $T\to 0$.

\item{\bf For both weak interactions and weak disorder} we still find a superconductor with the same $T_c$ and gap magnitude as in the absence of disorder.  When the disorder is sufficiently weak that $\ell \gg \xi_{0}$, this result is obvious.  However, for the case $\xi_0 \gg \ell \gg k_F^{-1}$, the result is highly non-trivial.  If we were to ignore the effects of disorder in computing  the quasi-particle scattering rate $1/\tau$  either just above $T_c$ or even below $T_c$, we would be off by a parametrically large factor $\tau_0/\tau \sim (k_F\xi_0)(\xi_0/\ell)$.  Indeed if we were to compute the zero temperature superfluid stiffness ignoring the effects of disorder we would be off by a factor of $(\xi_0/\ell)$ from the true value.  But by the miracle of ``Anderson's theorem'' - which is analogous to the cancellations in the ETh results discussed above - if we computed $T_c$ totally ignoring the effect of disorder on the electron propagator, we would get precisely the correct mean-field value.  Moreover, while fluctuation effects are enhanced by disorder, so long as $d>2$ the Ginzburg parameter $g =[ \rho(E_F) \Delta \xi^d]^{-1} \sim (k_F\ell)^{-d/2} (k_F\xi_0)^{(d-2)/2}$, still vanishes exponentially as $U\to 0$, meaning that the mean-field estimate of $T_c$ remans asymptotically exact.  (Recall that in a dirty superconductor, $\xi \sim \sqrt{\xi_0\ell}$.)
\end{itemize}

One other observation is worth making.  It is possible to define a limit in which the Migdal-Eliashberg theory for the electron-phonon problem is exact, regardless of the strength of the electron-phonon coupling or the degree of retardation.  Here we consider introducing $N^2$ flavors of  phonons and $N\times M$ flavors of fermions in a $O(N)\times O(M)$ symmetric manner, in which the electron-phonon coupling has the form
\begin{equation}
H_{el-ph} =\frac {\alpha}{[NM]^{1/4}} \sum_{\vec R} \psi^\dagger_{a,\alpha}(\vec R) X^{\alpha,\alpha^\prime}(\vec R)  \psi_{a,\alpha^\prime}(\vec R)
\end{equation}
where the sum over $\alpha$ and $\alpha^\prime = 1 - N$  and $a=1 - M$ is implicit.  In the limit $N\to \infty$ and $M\to \infty$ with $N/M = q$, the Migdal approximation (and correspondingly the ETh below $T_c$) is exact.  (In the case $q \gg 1$, where there are  many more flavors of boson than of fermions, the renormalization of the phonon propagator can be ignored.  Conversely, for $q \ll 1$, the renormalization of the fermions propagator is parametrically small.)

It is not, of course, clear how much of the relevant physics is captured by this peculiar large $N$ limit.  One interesting route to take, however, would be to examine the $1/N$ corrections to this theory, and to explore the extent to which their importance is controlled by the Migdal parameter $\lambda (\Omega_0/E_F)$ rather than the value of $\lambda$ itself.

\acknowledgements
  We thank  B. Altshuler,  E. Berg, R. Combescot,  R. Fernandes, A. Finkelstein,  A. Klein, G. Kotliar, S. Lederer, L. Levitov, D. Maslov, A. Millis, V. Pokrovsky, N. Prokofiev, S. Raghu, M. Randeria, S. Sachdev, D. Scalapino, Y. Schattner, J. Schmalian, B. Svistunov,   E. Yuzbashyan, Y. Wang, Y. Wu, and J. Zaanen for useful discussions.   The work by  AVC was supported by the Office
of Basic Energy Sciences, U.S. Department of Energy, under award  DE-SC0014402.  SAK was supported, in part, by NSF grant \# DMR-1608055 at Stanford.  IE acknowledges support from the Harvard Quantum Initiative Postdoctoral Fellowship in Science and Engineering.

\bibliography{eliashberg90phonons}

\end{document}